\documentclass{article}
\usepackage{graphicx}  
\usepackage{amsmath}   
\usepackage[compress]{cite}
\usepackage{amssymb}   
\usepackage{bm} 
\usepackage{dcolumn}
\usepackage{color}
\usepackage{mathrsfs}
\usepackage{amsfonts}
\usepackage{varioref}
\RequirePackage[colorlinks,citecolor=blue,urlcolor=magenta,linkcolor=blue]{hyperref}
\input epsf
\def\EH{Einstein-Hilbert }
\def\LL{Lanczos-Lovelock }

\def\gr{general relativity}
\def\RN{Reissner-Nordstr\"{o}m }

\def\GB{Gauss-Bonnet }
\labelformat{section}{Section #1} 
\labelformat{subsection}{Section #1} 
\labelformat{subsubsection}{Section #1}
\labelformat{subsubsubsection}{Section #1}
\labelformat{equation}{Eq.~(#1)} 
\labelformat{figure}{Fig.~#1} 
\labelformat{subfigure}{Fig.~\thefigure#1} 
\labelformat{table}{Tab.~#1} 
\labelformat{appendix}{Appendix #1}
\title{Excavating black hole continuum spectrum: Possible signatures of scalar hairs and of higher dimensions}
\author{Indrani Banerjee
\footnote{tpib@iacs.res.in},
Sumanta Chakraborty
\footnote{sumantac.physics@gmail.com}
and Soumitra SenGupta
\footnote{tpssg@iacs.res.in}\\
{\small{Department of Theoretical Physics, Indian Association for the Cultivation of Science, Kolkata-700032, India}}}
\begin{document}
  
\maketitle
\begin{abstract}

Continuum spectrum from black hole accretion disc holds enormous information regarding the strong gravity regime around the black hole and hence about the nature of gravitational interaction in extreme situations. Since in such strong gravity regime the dynamics of gravity should be modified from the Einstein-Hilbert one, its effect should be imprinted on the continuum spectrum originating from the black hole accretion. To explore the effects of these alternative theories on the black hole continuum spectrum in an explicit manner, we have discussed three alternative gravitational models having their origin in three distinct paradigms --- (a) higher dimensions, (b) higher curvature gravity and (c) generalised Horndeski theories. All of them can have signatures sculptured on the black hole continuum spectrum, distinct from the standard general relativistic scenario. Interestingly all these models exhibit black hole solutions with tidal charge parameter which in these alternative gravity scenarios can become \emph{negative}, in sharp 
contrast with the Reissner-Nordstr\"{o}m black hole. Using the observational data of optical luminosity for eighty Palomer Green quasars we have illustrated that the difference between the theoretical estimates and the observational results gets minimized for negative values of the tidal charge parameter. As a quantitative estimate of this result we concentrate on several error estimators, including reduced $\chi^{2}$, Nash-Sutcliffe efficiency, index of agreement etc. Remarkably, all of them indicates a negative value of the tidal charge parameter, signaling the possibility of higher dimensions as well as scalar charge at play in those high gravity regimes.

\end{abstract}
\section{Introduction}\label{Accretion_Intro}

General Relativity has been the most successful theory till date in describing the spacetime structure around us. Even though it has passed various observational tests so far with flying colors, it has also failed in certain respects. The most prominent one being the galaxy rotation curves and accelerated expansion of the universe \cite{Clifton:2011jh,Perlmutter:1998np,Riess:1998cb}. There have been efforts to explain these by either introducing some additional matter fields, in which case the gravity theory is still given by \gr\ or, by invoking additional higher curvature correction terms to \gr\ \cite{Padmanabhan:2002ji,Copeland:2006wr,Dadhich:2010ca,Capozziello:2011et,Nojiri:2017ncd}. The second option where modifications are done to the Lagrangian of the gravity sector, are known as alternative theories of gravity. Setting up an alternative gravity theory as a viable option is not an easy task. The theory has to pass through several theoretical and observational hurdles --- the theory must be free from ghost modes, i.
e., should not allow superluminal propagating degrees of freedom, the theory must be consistent with solar system 
tests and should not give rise to a fifth force in the local physics, the theory must have something more to present, e.g., possible explanation of late time acceleration. These alternative theories may also have some completely different origin, such as the presence of extra spatial dimensions, resulting into four dimensional gravitational field equations different from that in Einstein gravity \cite{Will:2014kxa}. 

Among the various alternative theories there exist broadly three classifications: (a) The gravitational Lagrangian is modified by introduction of higher curvature terms, e.g., \LL gravity, $f(R)$ gravity etc. The \LL theory is inherently ghost free, leading to second order field equations, while $f(R)$ gravity models have ghost modes unless some specific conditions are being satisfied (see e.g., \cite{Lanczos:1932zz,Lovelock:1971yv,gravitation,Padmanabhan:2013xyr,Dadhich:2008df,Chakraborty:2014rga,Chakraborty:2014joa,
Chakraborty:2015wma,Chakraborty:2015kva,Lochan:2015bha,Nojiri:2010wj,Nojiri:2003ft,Nojiri:2006gh,Capozziello:2006dj,Bahamonde:2016wmz,Chakraborty:2016ydo}); (b) Modified gravitational dynamics due to existence of extra dimensions, e.g., bulk Weyl tensor contributes additional terms to the effective lower dimensional gravitational field equations \cite{Shiromizu:1999wj,Dadhich:2000am,Harko:2004ui,
Carames:2012gr,Kobayashi:2006jw,Shiromizu:2002qr,Haghani:2012zq,Borzou:2009gn,Chakraborty:2014xla,Chakraborty:2015bja,
Chakraborty:2015taq,Chakraborty:2016gpg} and, finally (c) Scalar-tensor theories of gravity, which was first introduced as the Brans-Dicke theory and emerged in recent times in a more general context as Horndeski theories \cite{Horndeski:1974wa,Sotiriou:2013qea,Babichev:2016rlq,Brans:1961sx,Herdeiro:2015waa,Gleyzes:2014dya,Langlois:2015skt,
Deffayet:2009wt,VanAcoleyen:2011mj,MuellerHoissen:1989yv,Parattu:2016trq,Babichev:2012re,Kobayashi:2014eva,Babichev:2015rva,
Charmousis:2015txa,Bhattacharya:2016naa,Anabalon:2013oea,Cisterna:2014nua}. This alike Lovelock models have higher powers of second derivatives in the Lagrangian, while the field equations are still of second order. In particular it can be generalized to include higher form fields as well \cite{Deffayet:2010zh,Hinterbichler:2010xn}. These models are interlinked among themselves, e.g., Kaluza-Klein reduction of Lovelock models from higher spacetime dimensions to four dimensions leads to the Horndeski theories \cite{VanAcoleyen:2011mj,Charmousis:2014mia,Charmousis:2015txa}.

The best place to test these alternative theories is in the near horizon regime of a supermassive black hole, since in that regime the curvature will be high enough for these modifications leading to observable effects. The two most significant phenomenon associated with the near horizon geometry of supermassive black holes correspond to strong gravitational lensing and structure of accretion disc. While, the strong lensing will become an observable only a few years later, the spectrum of the accretion disc around black holes can be easily extracted from various X-ray telescopes as well as Very Long Baseline Interferometers (in short VLBI) to a good accuracy and hence can be used to find out the possible presence of alternative gravity theories \cite{Virbhadra:1999nm,Bozza:2001xd,Bozza:2009yw,Sahu:2015dea,Bhadra:2003zs,Ghosh:2010uw,Ayzenberg:2017ufk,Psaltis:2008bb,
Bambi:2015kza,Abramowicz:2011xu,Zhang:2016ach,Fish:2009va,Jusufi:2017vew,Shchigolev:2016gro,Tsukamoto:2016jzh} (see also \cite{Andriot:2017oaz}). We have already discussed the strong gravitational lensing in alternative gravity theories in an earlier work \cite{Chakraborty:2016lxo}. While in this work our main aim will be to understand the impact of alternative gravity theories on the electromagnetic observations from the accretion disc around supermassive black holes. We will chiefly concentrate on the continuum spectrum from the black hole accretion disc as it is highly sensitive to the geometric properties of the background metric. 

It has been well established that Active Galactic Nuclei (henceforth referred to as AGN), which are the most luminous persistent sources of electromagnetic radiation in the universe are accretion powered \cite{LyndenBell:1969yx,Kazanas:2012sk}. Residing at the center of most massive galaxies \cite{LyndenBell:1969yx,Lynden_Rees:1971,Rees:1984si}, AGNs emit copious amount of radiation in several bands of electromagnetic spectrum from the radio to the gamma rays \cite{Sanders:1988rz,Neugebauer:1987,Baskin:2004wn,Scott:2004sv,Brandt:1999cm}. The optical/UV emission of the AGN spectral energy density (referred to as SED) is believed to be emanated from the geometrically thin and optically thick Keplerian disk generally extending in the intermediate and larger distances from the black hole \cite{Pringle_Rees:1972,Pringle:1981ds,Shakura:1972te,Shakura:1976xk,Abramowicz:1988sp,Prendergast_Burbidge:1968,Afshordi_2003}. This emission resembles a multicolored black body spectrum peaking in the optical/UV/EUV range depending on the mass of the black hole.
Since the temperature of the accretion disk varies as $T \propto M_{\rm BH}^{-1/4}$ \cite{Frank:2002}, where $M_{\rm BH}$ represents the mass of the black hole, the blackbody spectrum for AGN disks generally peaks in the UV \cite{Sanders:1989tu}. Gravitational binding energy released during the process of accretion produces a luminosity  $\sim 10^{44-47} \rm~ erg s^{-1}$ \cite{Maccacaro:1991,dellaceca:1991}, while the inner accretion disk has a temperature of $10^5-10^6$ K \cite{Reynolds:2013rva,Blackburne:2011,Jimenez-Vicente:2014lta}.

The spectrum from the thin accretion disk around a black hole has been analytically computed in the Newtonian approximation \cite{Shakura:1972te,Shakura:1976xk} as well as in the presence of general relativity \cite{Thorne:1974ve,Page:1974he,Abramowicz_2010,Novikov_Thorne_1973,Riffert_Herold_1995,Davis:2010uq}. In this work we compute the emission from the Keplerian disk around a set of eighty Palomar Green (PG) Quasars \cite{Schmidt:1983hr,Davis:2010uq} assuming that the background metric is given by the alternative theories and contrast the results with the outcomes from \gr \cite{Blaschke:2016uyo,Stuchlik:2008fy,Schee:2008fc}. The bolometric luminosity of these quasars is determined based on optical \cite{Neugebauer:1987}, UV \cite{Baskin:2004wn}, far-UV \cite{Scott:2004sv}, and soft X-ray \cite{Brandt:1999cm} observations. The observational values of the optical luminosities are used to compute the accretion rates and these quantities are quoted in \cite{Davis:2010uq}. The masses of most of these PG quasars are estimated from the characteristic velocities and radii of the broad line region (known as BLR) and from high-quality spectroscopy of the H$\beta$ region given by \cite{Boroson:1992cf}, while the mass of the remaining quasars are obtained from the $M-\sigma$ relation 
\cite{Ferrarese:2000se,Gebhardt:2000fk,Dasyra:2006jz,Wolf:2008sm,Tremaine:2002js}.

In this work, we have tried to understand the continuum spectrum emitted by these quasars modulo existence of gravity theories beyond Einstein. To understand possible consequences of these modified theories on the electromagnetic emission from black hole accretion, we concentrate on three such alternative gravity models --- one of them corresponds to modifications of gravitational field equations due to existence of extra dimensions and the other modifies the gravitational field equations by incorporating higher curvature terms as, while the third one corresponds to Horndeski (or, generalised Horndeski) theories of gravity. The theoretically obtained optical luminosity is compared with the observed values to identify which of these alternative gravity models best reproduce the observations.

The paper is organized as follows: In \ref{Accretion_Alt} we introduce basics of these alternative gravity theories and then the formulations of accretion disc has been carried out in \ref{Accretion_general}. Subsequently, we apply this formalism to the black hole spacetimes under consideration in \ref{Accretion_Black}, while the observed data has been numerically analyzed in \ref{Accretion_Obs}. Finally, we conclude with a discussion on our results. 
\section{Alternative gravity theories: A brief introduction}\label{Accretion_Alt}

In this section we will provide a brief introduction to the alternative gravity theories to be discussed in this work. We will first provide the modifications brought to the gravitational field equations due to presence of extra dimensions as well as higher curvature terms in the gravitational action and subsequently shall introduce the Horndeski theories. 
\begin{itemize}

\item \textbf{Black hole in higher dimensional Einstein gravity:} In presence of extra spatial dimensions, it is legitimate to ask about the corresponding effect on the lower dimensional hypersurface we live in (called as the brane). Given the gravitational field equations in the bulk, which we take to be the Einstein's equations, one can obtain the corresponding equations on a lower dimensional hypersurface by projecting the appropriate geometric quantities. The projection on a timelike surface can be performed by using the projector $h^{a}_{b}=\delta ^{a}_{b}-n^{a}n_{b}$, such that $h^{a}_{b}n_{a}=0$. Then one can use the Gauss-Codazzi equation, which essentially projects the Riemann tensor on the timelike hypersurface to which $n_{c}$ is the normal. This leads to the intrinsic (or, in this case four dimensional) Riemann tensor on the hypersurface as well as extrinsic curvatures associated with the normal $n_{a}$. Contraction of the above relation leads to 
projections of the Ricci tensor \cite{Shiromizu:1999wj}. Finally combining all these curvatures together and using Israel junction conditions on the hypersurface along with $Z_{2}$ orbifold symmetry, the vacuum gravitational field equations on the brane become,
\begin{align}\label{Eq_Acc_01}
G_{\mu \nu}+E_{\mu \nu}=0;\qquad E_{\mu \nu}=~^{(5)}C_{\mu a \nu b}n^{a}n^{b}
\end{align}
Here, $E_{\mu \nu}$ is the additional contribution to the gravitational field equations, originating from the bulk Weyl tensor. In principle unless one knows about the structure of the bulk spacetime it is not possible to get a handle on this term and in which case the brane geometry gets automatically fixed. Thus in absence of any knowledge about the bulk spacetime, one must treat the $E_{\mu \nu}$ as a source term. Note that due to Bianchi identity one must have $\nabla_{\mu}E^{\mu}_{\nu}=0$, which along with \ref{Eq_Acc_01} leads to a system of closed equations to be solved to get the brane geometry. In particular one can derive static and spherically symmetric solutions associated with the above field equations, which has been carried out in \cite{Dadhich:2000am,Harko:2004ui}. The standard trick is to write down the bulk tensor $E_{\mu \nu}$ in terms of irreducible representations using a four velocity $u_{\mu}$, which leads to two unknown radial functions, the dark radiation $\mathcal{U}(r)$ and the 
dark pressure $\mathcal{P}(r)$ respectively. With the following choice for the equation of state: $2\mathcal{U}+\mathcal{P}=0$, one can solve the conservation relation to yield $\mathcal{U}(r)=\alpha (q/r^{4})$, where $\alpha$ is the fourth power of the ratio of four dimensional to five dimensional gravitational coupling constants. The corresponding spacetime structure is given by the following line element,
\begin{align}
ds^{2}=-\left(1-\frac{2M}{r}+\frac{q}{r^{2}}\right)dt^{2}+\left(1-\frac{2M}{r}+\frac{q}{r^{2}}\right)^{-1}dr^{2}+r^{2}d\Omega ^{2}
\end{align}
For $q>0$, this solution behaves as a \RN black hole, with an event horizon and a Cauchy horizon. While the case $q<0$ has no analogue in \gr\ and thus provides a true signature of additional spatial dimensions \cite{Dadhich:2000am}. 

\item \textbf{Black hole in Einstein-Gauss-Bonnet gravity:} So far, we have been considering the effect of higher spacetime dimensions to the gravitational dynamics on a $(1+3)$ dimensional brane. In this paragraph we will discuss another scenario, where due to the presence of a higher curvature term in the action the gravitational field equations are different from the Einstein's equations. One of the most promising such higher curvature scenario corresponds to Einstein-Gauss-Bonnet gravity for which the gravitational Lagrangian reads $R+\alpha \mathcal{G}$, where $\mathcal{G}=R_{\mu \rho \nu \sigma}R^{\mu \rho \nu \sigma}-4R_{\mu \nu}R^{\mu \nu}+R^{2}$ is the Gauss-Bonnet invariant and $\alpha$ is a positive constant of dimension $(\textrm{length})^{2}$ \cite{Boulware:1985wk}. In normal units, i.e., with $G=c=1$, the value of $\alpha$ is much small compared to unity. From a purely geometrical standpoint, the above theory is interesting in its own right, since even though the Lagrangian is higher order in 
the Riemann curvature, it still has only second order field equations. However the \GB term $\mathcal{G}$, corresponds to a total derivative in four spacetime dimensions. Thus in order to capture some non-trivial effects of the \GB term one must investigate the gravity theory in higher spacetime dimensions. Thus alike the previous situation, in this case as well we will be interested in four dimensional spherically symmetric solution admitting black holes, but originating from this higher curvature theory in higher dimensions. One such exact solution has been derived in \cite{Maeda:2006hj} and is known as the Maeda-Dadhich solution, which reads,
\begin{align}
ds^{2}=-f(r)dt^{2}+\frac{dr^{2}}{f(r)}+r^{2}d\Omega _{2}^{2}+\gamma _{AB}dx^{A}dx^{B}
\end{align}
The above $D$ dimensional spacetime has a product structure and factorizes in a spherically symmetric four dimensional spacetime and a $(D-4)$ dimensional Einstein space. Surprisingly, this solution has \emph{no} general relativistic limit and the function $f(r)$ in the limit of small \GB parameter $\alpha$ has the following form \cite{Bhattacharya:2016naa}
\begin{align}
f(r)=1-\frac{(D-4)\sqrt{\alpha}~m}{r}-\frac{(D-4)\alpha |\tilde{q}|}{r^{2}}
\end{align}
where, the effect of cosmological constant in the local physics has been neglected. Given this form of $f(r)$, one can identify the ADM mass and Newton's constant as: $2GM=(D-4)\sqrt{\alpha}~m$ and define a dimensionless constant $q=(D-4)|\tilde{q}|/m^{2}$ such that the geometry associated with the spherically symmetric brane (defined by $x^{A}=\textrm{constant}$) becomes (in the $G=1$ unit),
\begin{align}
ds^{2}=-\left(1-\frac{2M}{r}-\frac{4M^{2}q}{r^{2}} \right)dt^{2}+\left(1-\frac{2M}{r}-\frac{4M^{2}~q}{r^{2}} \right)^{-1}dr^{2}+r^{2}d\Omega ^{2}
\end{align}
Here the effect of extra dimensions as well as Gauss-Bonnet parameter is appearing through the tidal charge parameter $q$. The effect of \GB parameter is hidden in the definition of $q$ in terms of more primitive $\tilde{q}$, but the effect of higher dimensions can be immediately seen through the appearance of $q$ with a \emph{negative} sign. This is opposite to the positive signature of the electromagnetic term in the \RN black hole. This noteworthy feature of the solution signals the gravitational origin of this charge term, pointing towards existence of higher dimensions. Note that a similar result exists for the previous scenario as well, even though the origin of the charge term was completely different. In the first case it is due to the bulk Weyl tensor, while here it is due to the presence of Gauss-Bonnet term in the gravitational action.  

\item \textbf{Black hole in generalised Horndeski (scalar-vector-tensor) theory:} Having described purely gravitational modifications to the field equations, let us discuss another possibility, through non-minimal coupling of gravity with a scalar field. The model we will discuss regarding non-minimal scalar coupling corresponds to a generalisation of Horndeski theories. In general, Horndeski theories refer to the most general scalar-tensor theories with non-minimal couplings and can have various higher powers of derivatives of the scalar present in the action, even though the field equations are still second order \cite{Horndeski:1974wa,Sotiriou:2013qea,Babichev:2016rlq,Langlois:2015skt,Deffayet:2009wt}. 
While here we will also include an additional gauge field and shall couple it to the scalar sector non-minimally. This particular model has been explored earlier in detail 
\cite{Babichev:2013cya,Cisterna:2014nua,Babichev:2015rva}. However due to complicated nature of field equations it is difficult to discuss these theories with most generality. Despite the above, one can still consider a subset of the same, leading to simpler field equations. We will discuss one such scenario, where one couples the Maxwell's field to gravity as well as to a scalar field, which itself is non-minimally coupled with gravity. The corresponding action for the complete system, including gravity, scalar and gauge field takes the following form \cite{Babichev:2015rva}
\begin{align}
\mathcal{A}=\int d^{4}x~\sqrt{-g}\Bigg[\frac{R}{16\pi}-\frac{1}{4}F_{\mu \nu}F^{\mu \nu}
&+\beta G^{\mu \nu}\nabla_ {\mu}\phi \nabla_ {\nu}\phi-\eta \partial _{\mu}\phi \partial ^{\mu}\phi 
\nonumber
\\
&-\frac{\gamma}{2} \left(F_{\mu \sigma}F_{\nu}^{~\sigma}-\frac{1}{4}g_{\mu \nu}F_{\alpha \beta}F^{\alpha \beta}\right)\nabla ^{\mu}\phi \nabla ^{\nu}\phi \Bigg]
\end{align}
Here the action for pure gravity is taken to be the \EH term, for the gauge field it is the $-(1/4)F_{\mu \nu}F^{\mu \nu}$ term and finally for the scalar one has the canonical kinetic term. However in addition one has two more pieces --- (a) non-minimal coupling of gravity with scalar field through Einstein tensor and (b) coupling of the stress tensor of the gauge field to the scalar sector. These two pieces come with arbitrary dimensionful coefficients $\beta$ and $\gamma$ respectively. Even though the field equations in this simplified setting as well are complicated, one can use the additional symmetry $\phi \rightarrow \phi +\textrm{constant}$ to derive a conserved Noether current. Further imposing spherical symmetry to this problem simplifies the field equations considerably to obtain exact solutions. In the case with $\eta=0$, i.e., in absence of any canonical kinetic term for $\phi$ one obtains the following spherically symmetric solution \cite{Babichev:2015rva}
\begin{align}
ds^{2}=-\left(1-\frac{2M}{r}+\frac{\gamma(Q^{2}+P^{2})}{4\beta r^{2}}\right)dt^{2}+\left(1-\frac{2M}{r}+\frac{\gamma(Q^{2}+P^{2})}{4\beta r^{2}}\right)^{-1}dr^{2}+r^{2}d\Omega ^{2}
\end{align}
where the charges associated with the gauge field can be obtained from $F_{tr}=Q/r^{2}$ and $F_{\theta\phi}=P\sin \theta$. Further the scalar field (or, the Galileon field) present in this model takes the following form \cite{Babichev:2015rva}
\begin{align}
\phi(r)=qt+\psi(r);\qquad \psi'(r)^{2}=\frac{\frac{2M}{r}-\frac{\gamma(Q^{2}+P^{2})}{4\beta r^{2}}}{\left(1-\frac{2M}{r}+\frac{\gamma(Q^{2}+P^{2})}{4\beta r^{2}}\right)^{2}}q^{2}
\end{align}
Here the additional constant $q$ appearing in the solution for the scalar field is related to the coefficient of the non-minimal coupling between Galileon and gravity, as $q^{2}=1/\beta$. Thus one must have $\beta >0$ to ensure a real solution for $q$. Further the gauge field as well as the scalar field with the positive branch of the above equation for $\psi(r)$ is regular at the event horizon. At this stage one has no conditions on the coupling between the gauge field and the Galileon, thus unlike the higher dimensional scenario, the sign of $1/r^{2}$ term can have either signs. 

\end{itemize}
Note that in all the three distinct scenarios, the modifications to the gravity sector is being represented by the term $q/r^{2}$, which for the case $q>0$ corresponds to standard \RN like scenario. However the case $q<0$ is completely unique and originates solely due to the modifications to the Einstein's equations. This suggests some novel understanding regarding our universe in the strong gravity regime, with possibility of signatures of higher dimensions as well as higher curvature gravity along with scalar hairs. This also helps to envisage the key motivation behind this work, namely whether one can use the strong gravity regime near a supermassive black hole to extract information about the nature of gravity at such energy scales. To explore this idea further, in the later sections we will study the continuum spectrum originating from the accretion disc around supermassive black holes and shall witness what the spectrum has to tell about black hole hairs and hence about the fundamental structure of our 
spacetime.
\section{Accretion disc in a static and spherically symmetric spacetime: A general analysis}\label{Accretion_general}

In this section we will analyze the characteristics of electromagnetic emission from an accretion disc in a general static and spherically symmetric spacetime and hence the associated observables. The results derived in this context will be used extensively in the later sections to derive the observables associated with accretion disc for the alternative gravity theories introduced in \ref{Accretion_Alt}. For generality and keeping future applications in mind we will work out the details of the observables associated with the accretion process onto a black hole represented by the most general static and spherically symmetric metric ansatz in four spacetime dimensions, which reads,
\begin{equation}\label{Acc_Eq_01}
ds^{2}=-f(r)dt^{2}+\frac{dr^{2}}{g(r)}+r^{2}d\Omega ^{2}
\end{equation}
where, $f(r)$ and $g(r)$ are arbitrary functions of the radial coordinates and is dependent on the specific gravity theory under consideration. Due to spherical symmetry, we assume the disc to lie on the equatorial plane, i.e., $\theta =\pi/2$. The spacetime being static and spherically symmetric there exist two Killing vectors $\partial/\partial t$ and $\partial/\partial \phi$ respectively. This suggests existence of two conserved quantities, namely the energy $E$ and the angular momentum $L$ respectively. In the case of accretion disc, one is primarily interested in the structure of circular geodesics of a massive particle (characterized by $r=\textrm{constant}=r_{c}$), which for this metric structure has already been studied in \cite{Chakraborty:2015vla} and the expressions for energy and angular momentum read,
\begin{align}
E_{c}&=\sqrt{\frac{2f_{c}^{2}}{2f_{c}-r_{c}f_{c}'}}
\label{Acc_Eq_02a}
\\
L_{c}&=\sqrt{\frac{r_{c}^{3}f_{c}'}{2f_{c}-r_{c}f_{c}'}}
\label{Acc_Eq_02b}
\end{align}
where `prime' denotes differentiation with respect to the radial coordinate along with $f_{c}$ and $f'_{c}$ referring to values of $f(r)$ and $f'(r)$ at $r=r_{c}$ respectively. From these expressions it is clear that both the energy and angular momentum associated with the circular geodesic can be specified solely in terms of the radius of the circular orbit $r_{c}$. Basically given the geodesic equations, one imposes the following conditions $dr/d\phi=0=d^{2}r/d\phi ^{2}$ on them to get the energy and angular momentum of circular orbits. Using the geodesic equations and expressions from \ref{Acc_Eq_02a} and \ref{Acc_Eq_02b}, one obtains the angular velocity of the particle moving on the circular geodesic to be,
\begin{equation}\label{Acc_Eq_03}
\Omega \equiv\frac{d\phi}{dt}=\frac{L_{c}}{E_{c}}\frac{f_{c}}{r_{c}^{2}}\equiv \frac{\sqrt{M}}{r_{c}^{3/2}}\frac{1}{\mathcal{B}}
\end{equation}
where, $f_{c}$ denotes the value of $f(r)$ at the circular orbit radius $r=r_{c}$. Here we have introduced the quantity $\mathcal{B}$, which has the following form,
\begin{equation}\label{Acc_Eq_04}
\mathcal{B}=\sqrt{\frac{2M}{r_{c}^{2}f_{c}'}}
\end{equation}
Note that in the Schwarzschild spacetime $f'(r)=2M/r^{2}$ and hence, $\mathcal{B}=1$ \cite{Novikov_Thorne_1973}. Given the angular velocity, it is possible to get the corresponding linear velocity by introducing appropriate orthonormal frames. In this particular case, the relevant orthonormal frames are, $e^{(\phi)}_{i}=(0,0,0,r)$ as well as $e^{(t)}_{i}=(\sqrt{f(r)},0,0,0)$. Thus one can obtain the corresponding linear velocity of the particle traveling on a circular geodesic as,
\begin{equation}\label{Acc_Eq_05}
v^{(\phi)}=\frac{u^{i}e^{(\phi)}_{i}}{u^{j}e^{(t)}_{j}}=\frac{r}{\sqrt{f}}\Omega\equiv \sqrt{\frac{M}{r_{c}}}\frac{1}{\mathcal{B}\sqrt{\mathcal{D}}}
\end{equation}
where, the quantity $\mathcal{B}$ has been defined earlier and here we have introduced one more object, $\mathcal{D}=f_{c}$. Since the particle is in a circular orbit one can easily convince himself that the three velocity of the particle is $\mathbf{v}=(0,0,v^{(\phi)})$. Moreover, in the asymptotic limit (viz. $r\rightarrow \infty$) it reduces to $r\dot{\phi}$, thus setting $v^{(\phi)}=1$ yields the photon circular orbit, following standard expectation. Having obtained the linear velocity, one can compute the Lorentz factor corresponding this velocity, which reads,
\begin{equation}\label{Acc_Eq_06}
\gamma =\left[1-\left(v^{(\phi)}\right)^{2}\right]^{-1/2}\equiv \frac{\mathcal{B}\sqrt{\mathcal{D}}}{\sqrt{\mathcal{C}}}
\end{equation}
where the quantity $\mathcal{C}$ introduced above has the following expression,
\begin{align}\label{Acc_Eq_07}
\mathcal{C}=\frac{M\left(2f_{c}-r_{c}f_{c}'\right)}{r_{c}^{2}f_{c}'}
\end{align}
Here again $f_{c}'$ denotes derivative of $f(r)$ evaluated at the circular orbit, located at $r=r_{c}$. Using these two informations one can construct the four-velocity of the particle moving on the circular geodesic along the following lines. The normalization of the four velocity demands, $u^{t}=N\gamma$, which from \ref{Acc_Eq_05} helps to determine $u^{\phi}=\sqrt{f}N\gamma v^{(\phi)}/r$. This in turn determines the normalization to be $N=1/\sqrt{f}$. Thus the complete structure of the four velocity becomes, 
\begin{align}\label{Acc_Eq_08}
u^{\mu}=\frac{\gamma}{\sqrt{\mathcal{D}}}\left(1,0,0,\frac{\sqrt{\mathcal{D}}}{r_{c}}v^{(\phi)}\right)=
\frac{\mathcal{B}}{\sqrt{C}}\left[\left(\frac{\partial}{\partial t}\right)^{\mu}+\sqrt{\frac{M}{r_{c}^{3}}}\frac{1}{\mathcal{B}}
\left(\frac{\partial}{\partial \phi}\right)^{\mu}\right]
\end{align}
Given the four velocity, it is possible to compute the shear tensor associated with this congruence of circular geodesics on the equatorial plane, leading to $\sigma _{r\phi}=-(3/4)(\mathcal{D}/\mathcal{C})\sqrt{(M/r_{c}^{3})}$ as the only non-zero component. This expression will find its use later on. 

The other quantity of interest corresponds to marginally stable circular orbits, often denoted by $r_{\rm ms}$. This actually determines the location of the last stable circular orbit. After crossing this radius the particle has to fall inside the black hole. Thus $r_{\rm ms}$ provides the minimum radius after which (i.e., for $r<r_{\rm ms}$) the accretion disc structure gets destroyed. Computation of this radius can be done by noting the points of inflection of the effective potential $V_{\rm eff}(r)$, which corresponds to $V_{\rm eff}''(r)=0$ along with $V_{\rm eff}'(r)=0$. Use of both these conditions lead to the following algebraic equation 
\begin{equation}\label{Acc_Eq_09}
2rf(r)f''(r)-4rf(r)'^{2}+6f(r)f'(r)=0
\end{equation}
depending solely on the $g_{tt}$ component, $f(r)$ and its derivatives. For $f(r)=1-(2M/r)$, i.e., for Schwarzschild, one can solve the above equation, leading to $r_{\rm ms}=6M$, which will receive corrections in the alternative theories.

In order to proceed further and obtain the total flux out of the accretion disc in a closed form, we will have to make certain reasonable assumptions regarding the structure and properties of the accretion disc \cite{Novikov_Thorne_1973}. First of all, we assume that the central plane of the accretion disc coincides with the equatorial plane of the black hole. 
We assume the disc structure to be thin, i.e., $h(r)/r\ll1$. Moreover, the disc is assumed to be in a quasi-steady state, i.e., when averaged over a length scale $\sim \mathcal{O}(\textrm{height of the disc})$ and time scale $\sim \mathcal{O}(\textrm{time interval required for the accreting gas to travel a radial distance equivalent to the disc scale height})$, any turbulent phenomenon are supposed to be washed out, resulting in a steady state behaviour of physical quantities. Finally, one assumes that 
macroscopically the gas of the disc moves in almost circular orbits with an additional small radial velocity arising due to viscous stresses. Further, the velocity component in the vertical direction is assumed to be negligibly small. 

At this stage, one must write down the corresponding stress energy tensor associated with the fluid accreting around black holes. The stress energy tensor must involve four contributions --- (a) The standard energy-momentum tensor due to the geodesic flow $\rho _{0}u^{\alpha}u^{\beta}$; (b) Specific internal energy of the system $\Pi$; (c) Stress-tensor as measured in local rest frame of the accreting fluid $t^{\alpha \beta}$ and finally, (d) the energy flux $q^{\alpha}$. The corresponding expression for stress-energy tensor takes the following form,
\begin{align}\label{Acc_Eq_10}
T^{\mu \nu}\equiv \rho_{0}\left(1+\Pi\right)u^{\mu}u^{\nu}+t^{\mu \nu}+u^{\mu}q^{\nu}+q^{\mu}u^{\nu}
\end{align}
The flux of radiation going out of this system can be obtained by averaging the $(0,z)$ component of the above stress-energy tensor, which essentially corresponds to $\langle q^{z}\rangle$. The conservation of mass, along with the previous assumptions implies the rate of mass loss to be approximately constant and is denoted by $\dot{M}_{0}$. Subsequently, one can impose conservation of angular momentum, which reads, $\nabla _{\mu}[T^{\mu}_{\nu}(\partial /\partial \phi)^{\nu}]$. Finally the energy conservation demands, $u^{\nu}\nabla _{\mu}T^{\mu}_{\nu}=0$, which reads,
\begin{align}\label{Acc_Eq_11}
\rho_{0}\frac{d\Pi}{dt}+\nabla _{\alpha}q^{\alpha}+\sigma _{\alpha \beta}t^{\alpha \beta}+\frac{1}{3}\theta t^{\alpha}_{\alpha}+a^{\mu}q_{\mu}=0
\end{align}
Here, the first term depicts transfer of energy to internal forms, while $\nabla _{\alpha}q^{\alpha}$ denotes transportation of energy out of the accretion disc and the last term is a correction to the energy flow vector $q^{\alpha}$. The other two terms are the source terms: (a) $\sigma _{\alpha \beta}t^{\alpha \beta}$ denotes energy generation by viscous heating; (b) $\theta t^{\alpha}_{\alpha}$ represents energy generation due to compression while (c) $a^{\mu}q_{\mu}$ is related to the inertia of the flowing energy $\textbf{q}$. Using the facts that the accreting gas is in quasi-steady state as well as the nearly geodesic motion of the gas particles, one can obtain a simpler version of energy conservation. This result when coupled with the angular momentum conservation leads to the following expression for flux, 
\begin{equation}\label{Acc_Eq_12}
F=\frac{3\dot{M}_{0}M}{8\pi r^{3}}\frac{\mathcal{Q}}{\mathcal{B}\sqrt{\mathcal{C}}}
\end{equation}
where, $r$ is the radial distance and the quantity $\mathcal{Q}$ is defined as,
\begin{equation}\label{Acc_Eq_13}
\mathcal{Q}=\mathcal{L}-\frac{3}{2}\sqrt{\frac{M}{r}}\mathcal{I}\int _{r_{\rm ms}}^{r}dr~\frac{\mathcal{L}}{\mathcal{B}\mathcal{C}\mathcal{I}}\sqrt{\frac{M}{r^{3}}}
\end{equation}
Here $\mathcal{L}$ is defined in terms of the marginally stable circular orbit $r_{\rm ms}$ as,
\begin{equation}\label{Acc_Eq_14}
\mathcal{L}=\frac{1}{\sqrt{\mathcal{C}}}-\frac{L_{\rm ms}}{\sqrt{Mr}}
\end{equation}
where $L_{\rm ms}$ is the angular momentum associated with the marginally stable circular orbit. Further one must compute the following integral, leading to the expression for $\mathcal{I}$,
\begin{equation}\label{Acc_Eq_15}
\mathcal{I}=\exp\left[-\frac{3}{2}\int _{r}^{\infty}dr~\frac{M}{\mathcal{B}\mathcal{C}r^{2}} \right]
\end{equation}
It is clear from the above discussion that the flux from the accretion disk varies with the radial distance from the black hole. The disk emits locally as a black body, such that the radiated energy follows a Planck distribution with the effective temperature at a particular radial distance given by, $T_{\rm eff}={(F/\sigma•)}^{1/4}$. The luminosity from the disk is obtained by integrating the Planck function over the disk surface,
\begin{equation}\label{Acc_Eq_16}
L_\nu=8\pi^2 \cos i R_g^2 \int_{R_{\rm in}}^{R_{\rm out}} B_{\nu}(T_{\rm eff}(r))r dr,
\end{equation}
where, $L_\nu$ is the luminosity from the disk at frequency $\nu$ assumed to be emitted over the $4\pi$ solid angle, $i$ is the inclination of the disk to the line of sight and $B_{\nu}$ is the Planck function. The theoretically derived optical luminosity is obtained by, $L_{\rm opt}\equiv \nu L_{\nu}$ where the value of $\nu$ correspond to the wavelength of 4681\AA \cite{Davis:2010uq}. As nearly edge on systems are likely to be obscured, we assume $\cos i \sim 0.5-1$ for the systems under consideration. Following \cite{Davis:2010uq}, we adopt a typical value for $\cos i \sim 0.8$. We will discuss about this particular choice of the inclination angle in somewhat more detail in the later sections.

This completes our discussion on the accretion disc observables in terms of the metric elements in an arbitrary static and spherically symmetric spacetime. In particular, we have expressed the energy flux coming from the accretion disc in terms of the unknown functions $f(r)$, $g(r)$ and their derivatives. Thus we can use this result to derive the corresponding flux, had these accreting black holes be solutions of some alternative theories which can possibly lead to stringent constraints on these theories. We will next take up the job of computing the flux from accreting black holes in higher dimensional, higher curvature and generalized Horndeski theories for the scenarios detailed in \ref{Accretion_Alt}. 
\section{Accretion disc structure in alternative gravity theories}\label{Accretion_Black}

In this section we will explicitly discuss the structure as well as observables associated with accretion around supermassive black holes in the context of alternative theories elaborated in \ref{Accretion_Alt}. A quick look at all the black hole solutions would convince one that, these solutions even though have completely different origin have superficial mathematical similarity. In particular all of them can be written as,
\begin{equation}\label{Acc_Eq_17a}
f(r)=g(r)=1-\frac{2}{r}+\frac{4q}{r^{2}}
\end{equation}
Here we have introduced a new length scale $r_{\rm new}=r_{\rm old}/M$, which for notational convenience will be denoted by $r$ only. The parameter $q$ is a constant and dependent on various couplings and physical charges of these various models. For example, in the context of Einstein-Gauss-Bonnet gravity in higher spacetime dimensions, one have $q$ to be dependent on the \GB parameter $\alpha$ besides being negative. In the rest of the analysis we will keep the parameter $q$ arbitrary and shall let the continuum spectrum emitted by accretion disc around supermassive black holes to provide the necessary information regarding the same, which can possibly indicate to the nature of gravity at such high curvature regimes. Given the above metric elements one can compute the following quantities, necessary for flux computation, at some given circular orbit with radius $r$, as
\begin{equation}\label{Acc_Eq_17b}
\mathcal{B}=\left[1-\frac{4q}{r}\right]^{-1/2};\qquad \mathcal{C}=\frac{1-\frac{3}{r}+\frac{8q}{r^{2}}}{1-\frac{4q}{r}};\qquad \mathcal{D}=1-\frac{2}{r}+\frac{4q}{r^{2}}
\end{equation}
Use of these results to \ref{Acc_Eq_03}, \ref{Acc_Eq_06} and \ref{Acc_Eq_08} will lead to expressions for angular velocity, Lorentz factor and four velocity respectively as functions of $q$. Thus any geometrical objects associated with accretion flow can be determined as these three quantities in \ref{Acc_Eq_17b} are being used. 
With a knowledge of $\mathcal{B}$, $\mathcal{C}$, $\mathcal{D}$ and $f(r)$ one can work out $\mathcal{L}$, $\mathcal{I}$ and $\mathcal{Q}$ as discussed in the previous section and thus obtain the flux $F$ from the accretion disk. We use \ref{Acc_Eq_16} to derive the optical luminosity from the thin disk in the background of these alternative gravity theories.

The other important object corresponds to the location of marginally stable circular orbit, $r_{\rm ms}$, already discussed in \ref{Accretion_general}. This can be obtained by solving the following algebraic equation, 
\begin{equation}\label{Acc_Eq_18}
r^{3}-6r^{2}+36qr-64q^{2}=0
\end{equation}
derived using the form of $f(r)$ and $g(r)$ in \ref{Acc_Eq_09}, leading to,
\begin{align}\label{Acc_Eq_19}
r_{\rm ms}(q)=2&+2^{2/3}\Big(2+8q^{2}-9q+\sqrt{64q^{4}-36q^{3}+5q^{2}}\Big)^{1/3}
\nonumber
\\
&+2^{-2/3}\Big(2+8q^{2}-9q+\sqrt{64q^{4}-36q^{3}+5q^{2}}\Big)^{-1/3}\Big(4-12q\Big)
\end{align}
For $q=0$, one can immediately check that $r_{\rm ms}(0)=6$, demonstrating that we get back the Schwarzschild geometry correctly under appropriate limits. The flux generated by the accretion disc depends on the charge parameter $q$ non-trivially. At this stage we should mention that not all values of $q$ are allowed. For positive $q$, one must have $q\leq 0.25$, otherwise the event horizon would disappear and a naked singularity would be generated. On the other hand, for negative $q$ any value of the charge parameter is allowed. With this parameter space in mind the flux generated by viscous stresses within the accreting material takes the following form,
\begin{align}\label{Acc_Eq_20}
F(q)=\frac{3\dot{M}_{0}M}{8\pi r^{3}}&\frac{1-\frac{4q}{r}}{\sqrt{1-\frac{3}{r}+\frac{8q}{r^{2}}}}
\Bigg\lbrace \sqrt{\frac{1-\frac{4q}{r}}{1-\frac{3}{r}+\frac{8q}{r^{2}}}}
-\frac{L_{\rm ms}}{\sqrt{M^{2}r}}
\nonumber
\\
&-\frac{3}{2\sqrt{r}}\exp \left[-\frac{3}{2}\int _{r}^{\infty}\frac{d\bar{r}}{\bar{r}^{2}}
\frac{\left(1-\frac{4q}{\bar{r}}\right)^{3/2}}{1-\frac{3}{\bar{r}}+\frac{8q}{\bar{r}^{2}}} \right]
\int _{r_{\rm ms}}^{r}~\frac{d\bar{r}}{\bar{r}^{3/2}}
\left(\sqrt{\frac{1-\frac{4q^{2}}{\bar{r}}}{1-\frac{3}{\bar{r}}+\frac{8q^{2}}{\bar{r}^{2}}}}
-\frac{L_{\rm ms}}{\sqrt{M^{2}\bar{r}}}\right)
\nonumber
\\
&\times\exp \left[\frac{3}{2}\int _{\bar{r}}^{\infty}\frac{dr'}{r'^{2}} 
\frac{\left(1-\frac{4q}{r'}\right)^{3/2}}{1-\frac{3}{r'}+\frac{8q}{r'^{2}}} \right]
\frac{\left(1-\frac{4q^{2}}{\bar{r}}\right)^{3/2}}{1-\frac{3}{\bar{r}}+\frac{8q^{2}}{\bar{r}^{2}}}\Bigg\rbrace 
\end{align}
Having obtained the radiation flux from the accretion disc around a supermassive black hole, one can obtain the corresponding luminosity by assuming a Planck distribution for the radiated energy. 
\begin{figure}[t!]
\centering
\includegraphics[scale=1]{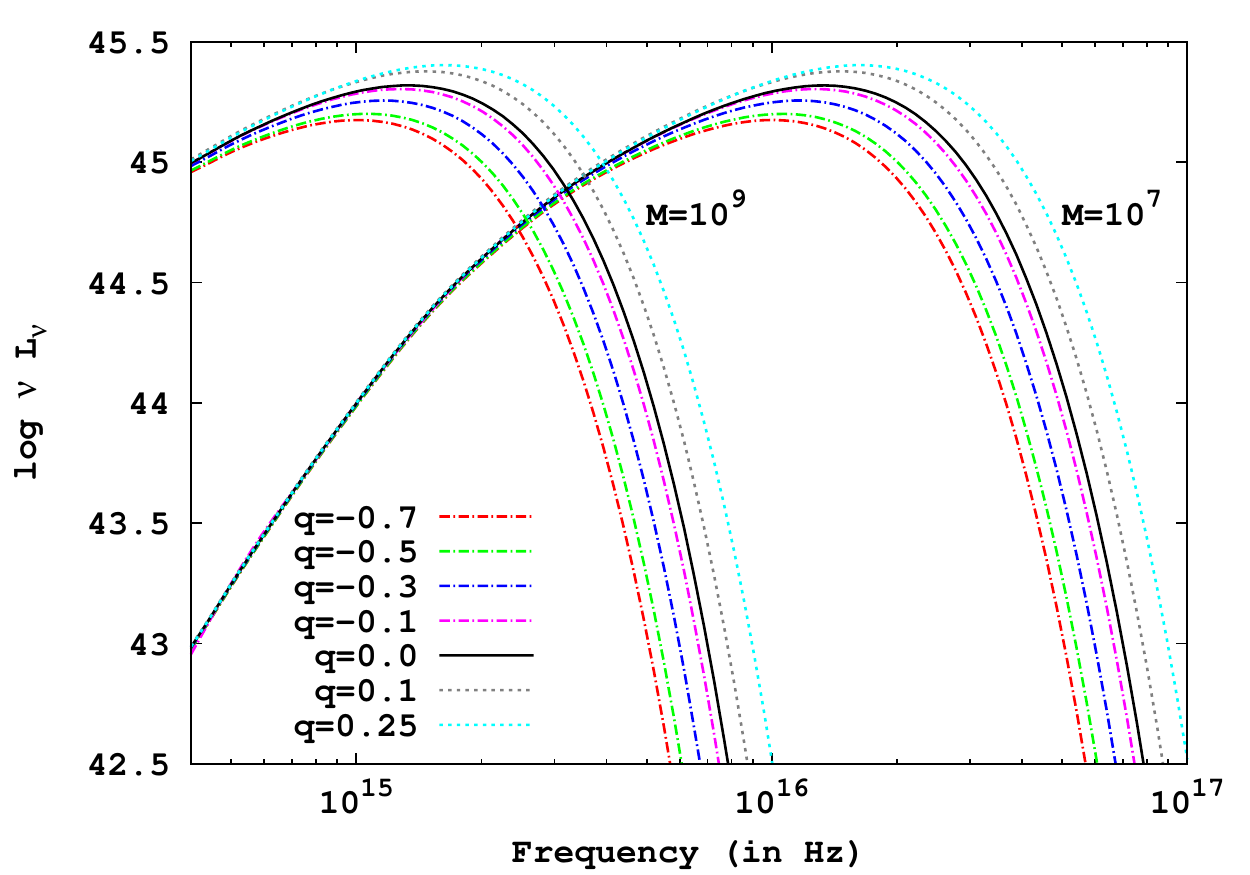}
\caption{The above figure illustrates the variation of the theoretically derived  luminosity from the accretion disc with frequency for two different masses of the supermassive black hole and for seven different values of $q$, both positive and negative. The case $q=0$ corresponds to the standard Schwarzschild scenario, which is depicted by the thick black line in the middle. Further, a clear distinction between positive and negative $q$ values is evident from the above figure, for positive values of $q$ the optical luminosity is larger than the Schwarzschild value, while for negative values of $q$ it is smaller. The accretion rate assumed is $1 M_{\odot}\textrm{yr}^{-1}$ and $\cos i$ is taken to be 0.8. See text for more discussions.}
\label{Fig_01}
\end{figure}
Since the luminosity depends on the charge parameter $q$, one can plot the luminosity against frequency for different choices of $q$. This will clearly demonstrate how the presence of $q$ affects the emission from the accretion disc. As anticipated, from \ref{Fig_01} it is clear that $q$ affects the optical luminosity in a \textit{non-trivial} manner. The modifications in the luminosity due to presence of $q$ is small at low frequencies, but the high energy behavior is significantly different compared to the Schwarzschild scenario. In particular the sign of $q$ plays a very important role in the departure of the luminosity from Schwarzschild counterpart, since the positive $q$ enhances luminosity at high energies, while negative charge parameter lowers the luminosity at high frequencies. Thus there is a distinctive signature of extra dimensions, which predict a negative value for $q$, on the continuum spectrum from black hole accretion disc. The same applies to black hole scalar hairs as well. In the next 
section we will scrutinize this surprising testbed for alternative theories yielding possible signatures of extra dimensions as well as nature of black hole hairs in detail.
\section{Numerical analysis: What does observations tell us?}\label{Accretion_Obs}

\begin{figure}[t!]
\centering
\vspace{-2.7cm}
\hbox{\hspace{-3.2em}
\includegraphics[scale=0.63]{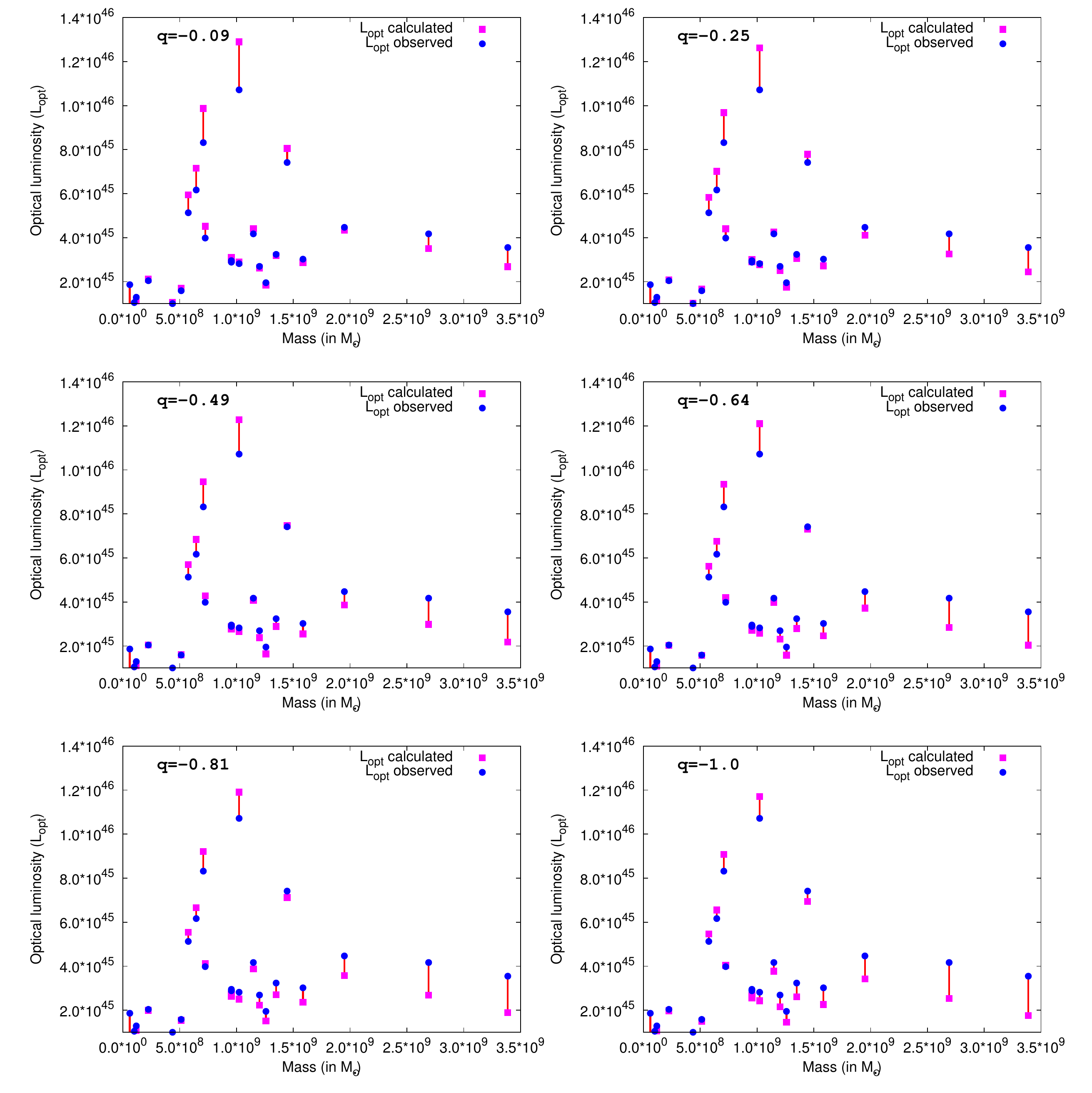}}
\caption{Theoretical optical luminosities along with the observed luminosity values for high mass PG quasars have been plotted with black hole mass $M$ for different negative values of the tidal charge parameter $q$. The differences between theoretical and observational values have also been depicted. It is quiet clear that some intermediate value of negative $q$ should minimize the difference between theoretical estimates and observed data. See text for more discussions.}
\label{Fig_L_M_nq}
\end{figure}
\begin{figure}[t!]
\centering
\vspace{-2.7cm}
\hbox{\hspace{-3.2em}
\includegraphics[scale=0.63]{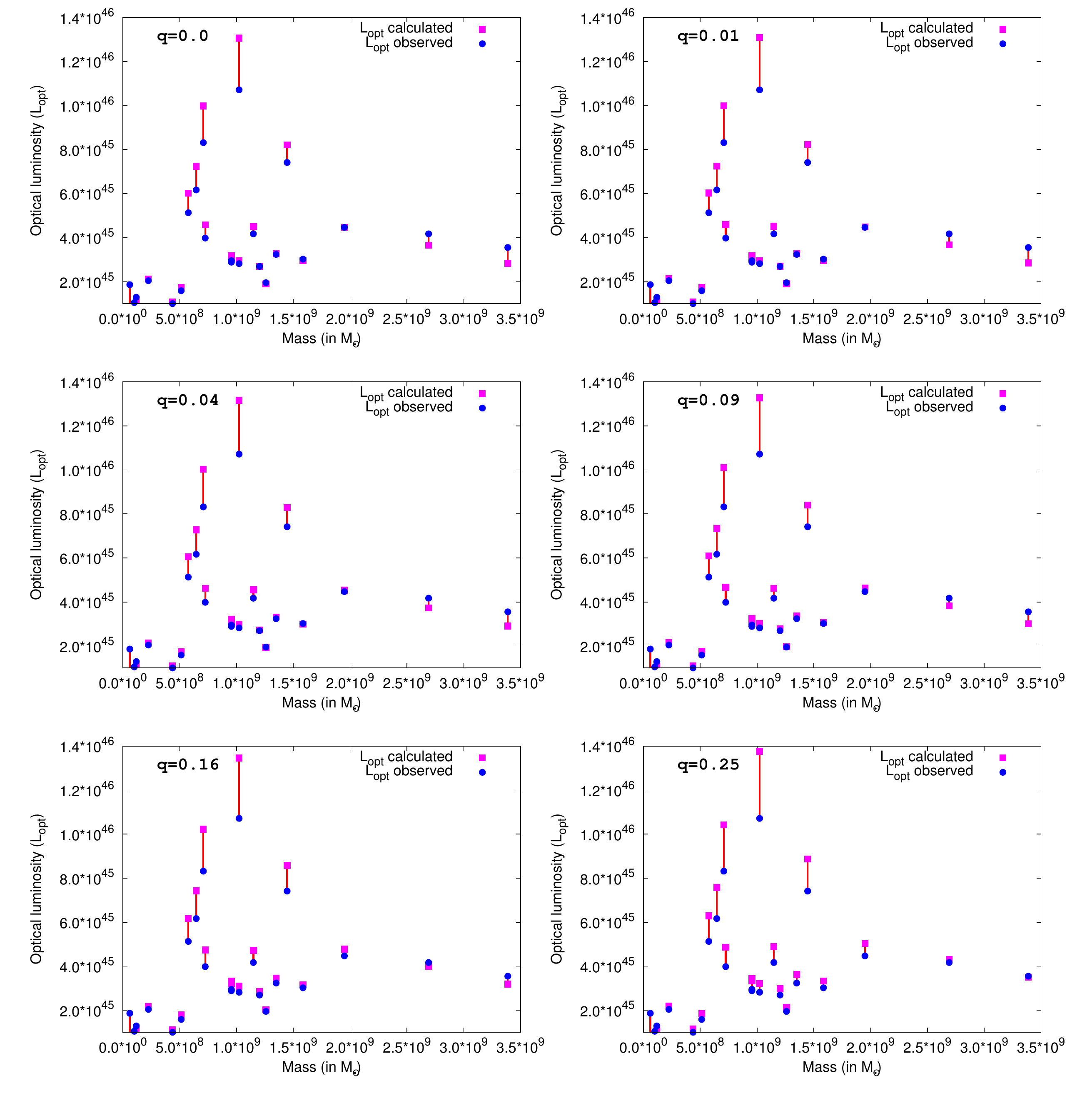}}
\caption{Theoretical optical luminosities associated with the electromagnetic radiation emanating from the accretion discs around the most massive black holes among the eighty PG quasars have been plotted against the black hole mass $M$ for different positive values of $q$. The same plot also depicts the corresponding observed values. The difference between the two values have been depicted in red lines. It is obvious that as one increases the value of $q$ the difference between the theoretical values and observed data also increases. See text for more discussions.}
\label{Fig_L_M_pq}
\end{figure}
\begin{figure}[t!]
\centering
\vspace{-2.7cm}
\hbox{\hspace{-3.2em}
\includegraphics[scale=0.63]{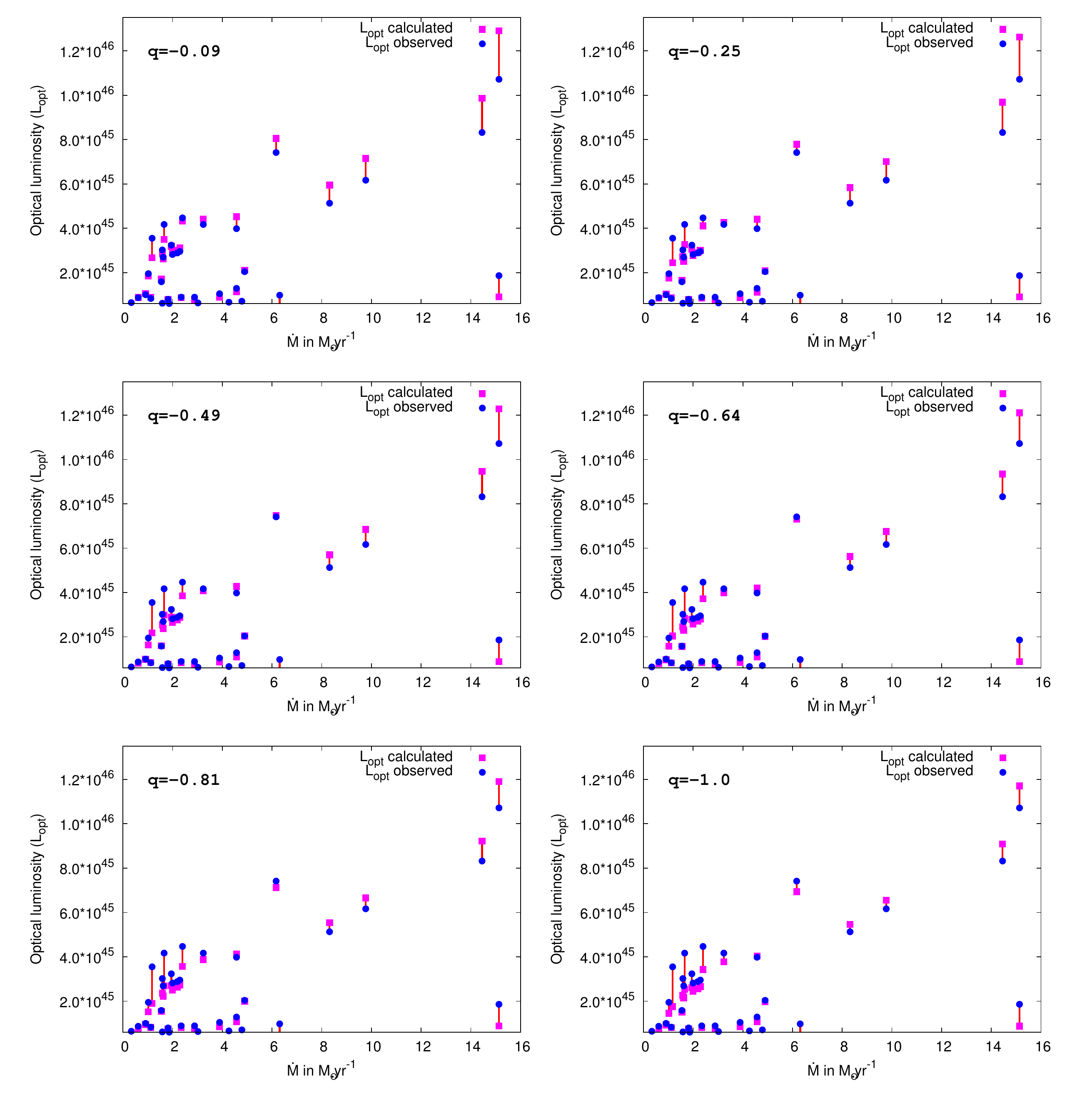}}
\caption{Theoretical optical luminosity for different negative values of $q$, along with observed luminosity values have been plotted for the high mass black holes among the eighty PG quasars against the accretion rate $\dot{M}$. The difference between theoretical and observational values have also been illustrated. In this case also it is clear that for some intermediate negative value for $q$ the difference between theoretical values and observed data are minimum.}
\label{Fig_L_dM_nq}
\end{figure}
\begin{figure}[t!]
\centering
\vspace{-2.7cm}
\hbox{\hspace{-3.2em}
\includegraphics[scale=0.63]{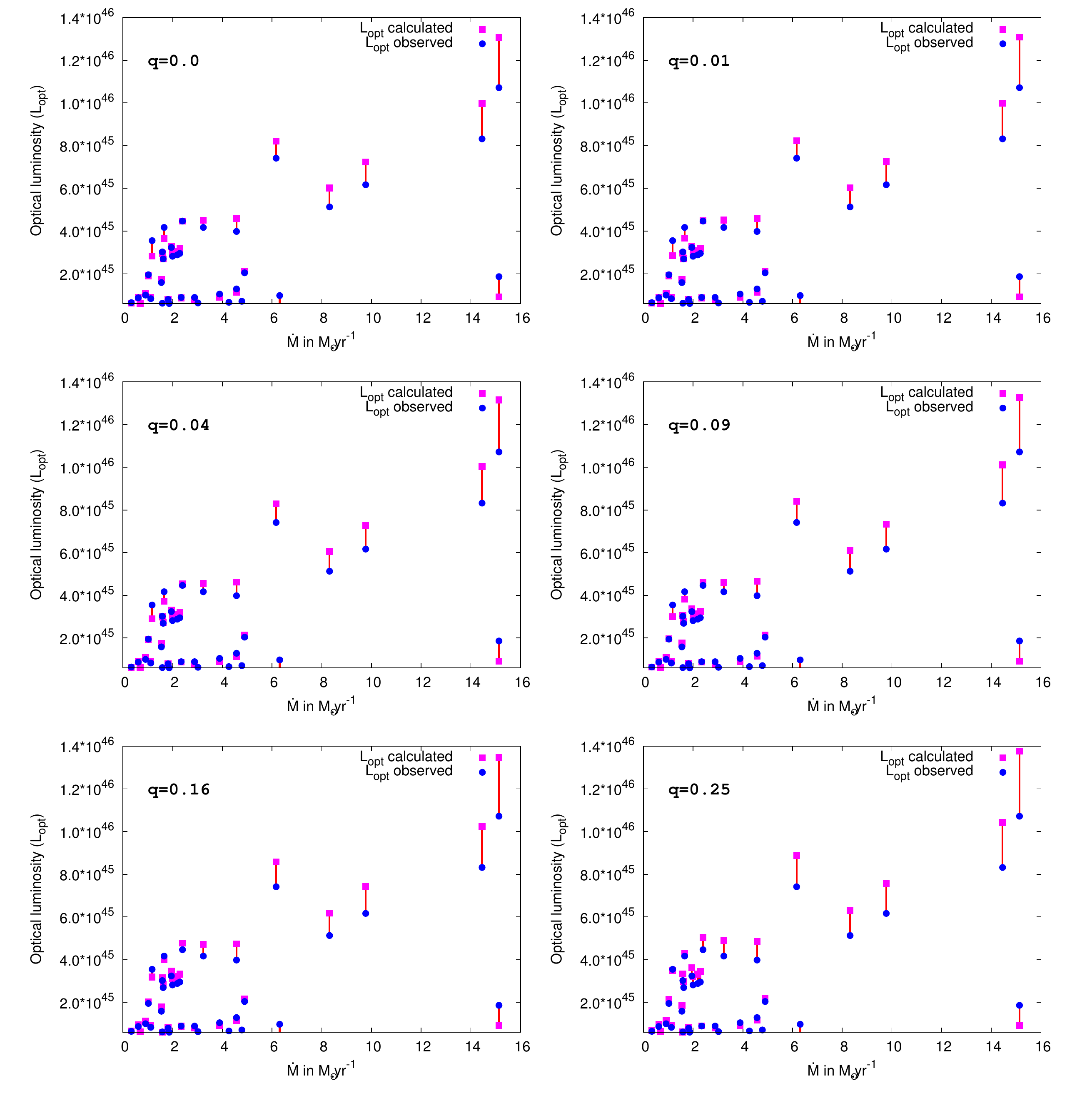}}
\caption{Theoretical optical luminosities for different positive values of $q$, along with the observed luminosity values have been plotted for high mass black holes among the eighty PG quasars against the mass accretion rate. The difference between theoretical and observational values can be clearly seen. The increase in the difference between theoretical values and observed data as $q$ is being increased, is also present here.}
\label{Fig_L_dM_pq}
\end{figure}
In this section we derive the theoretical values of optical luminosity for a sample of eighty PG quasars studied in \cite{Davis:2010uq} using the metric given by \ref{Acc_Eq_17a} as the background. 
The characteristic velocities and radii of the BLR as well as the high-quality spectroscopy of the H$\beta$ region \cite{Boroson:1992cf} have been used to constrain the masses of most of these PG quasars \cite{Davis:2010uq}. For the remaining quasars, the tight correlation between the mass $M$ and the stellar velocity dispersion $\sigma$ is exploited to estimate their masses \cite{Ferrarese:2000se,Gebhardt:2000fk}. There are thirteen such sources for which $\sigma$ is estimated from \cite{Dasyra:2006jz} and \cite{Wolf:2008sm} and the masses derived from the relation are given in \cite{Tremaine:2002js}. One can determine the bolometric luminosity using a number of observations, e.g., high quality optical \cite{Neugebauer:1987}, UV \cite{Baskin:2004wn}, far-UV \cite{Scott:2004sv}, and soft X-ray \cite{Brandt:1999cm}. The observational values of the optical luminosities and the accretion rates of these eighty PG Quasars are quoted in \cite{Davis:2010uq}. For every value of $q$ we compare the theoretically 
derived optical luminosity with the observed value for all the eighty PG quasars and calculate the reduced $\chi^2$ estimates, the Nash-Sutcliffe Efficiency, the index of agreement and the modified forms of the Nash-Sutcliffe efficiency and the index of agreement. This analysis enables us to predict the value of $q$ which is most favoured from observational considerations.
\begin{figure}[t!]
\centering
\includegraphics[scale=1]{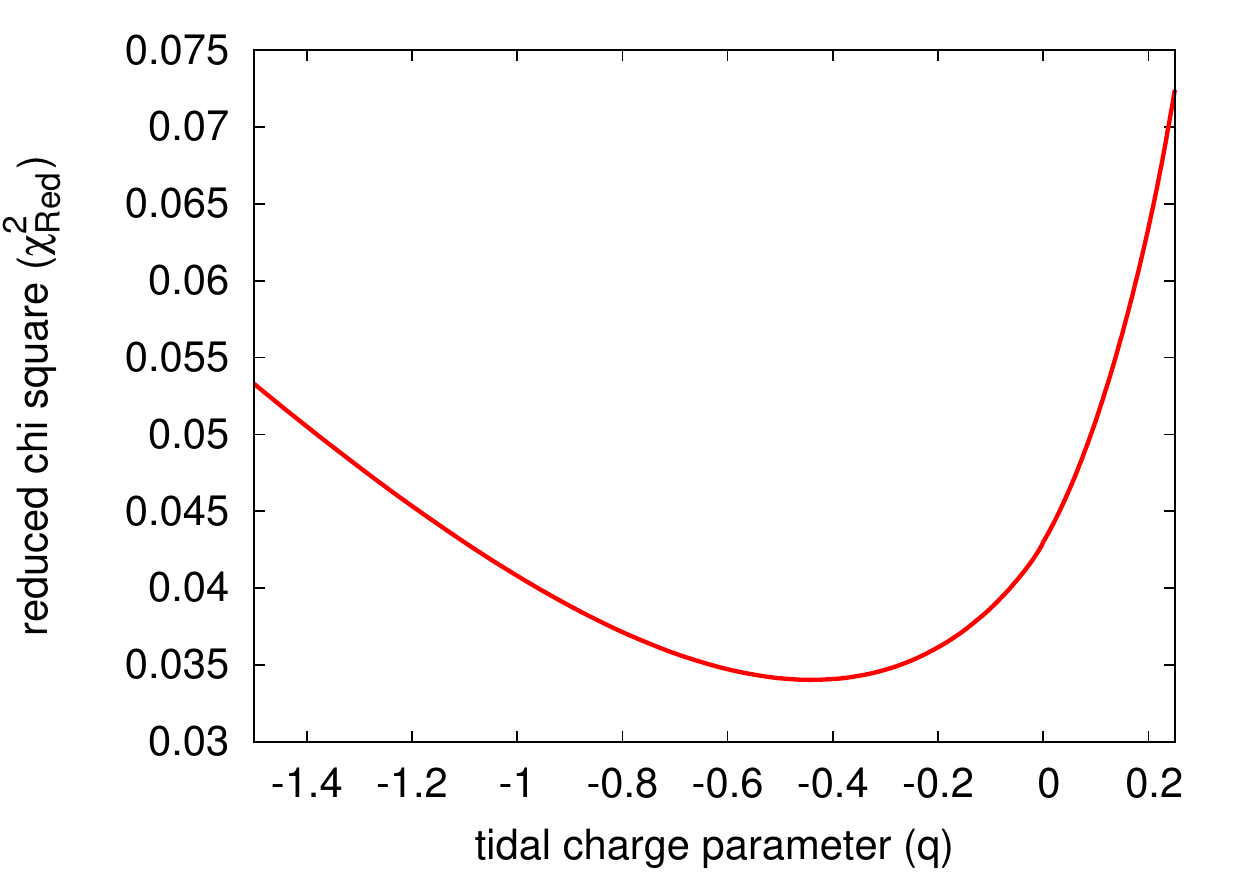}
\caption{The figure demonstrates variation of $\chi^{2}_{\rm Red}$ against the tidal charge parameter $q$ obtained using a sample of eighty supermassive black holes. Surprisingly, $\chi ^{2}_{\rm Red}$ attains minimum value for a negative value of $q\sim -0.4$, suggesting that the errors between theoretical estimates with observations are minimum when one considers either higher dimensional models or scalar hairs with a specific coupling to gauge fields. See text for more discussions.}
\label{Fig_chi}
\end{figure}

To understand the effect of tidal charge parameter on the electromagnetic spectrum emanating from the accretion disc in a qualitative manner, we plot the observed optical luminosity with the black hole mass and accretion rate along with theoretical estimates for the luminosity. The difference between the observed and theoretical values may indicate a preferred value for $q$. Keeping this in mind, in \ref{Fig_L_M_nq}, the theoretically derived optical luminosities and the corresponding observed values are plotted as a function of the mass of the quasars for various negative values of $q$. The figure illustrates that for intermediate and lower masses
the difference between the calculated and the observed luminosities is more for less negative values of $q$. On the other hand, towards the high mass end, this difference is more for more negative values of $q$. Thus, one can see from this figure that for the entire sample of eighty quasars the predicted theoretical luminosities is closest to the corresponding observed values for some intermediate negative value of $q$. \ref{Fig_L_M_pq} mimics \ref{Fig_L_M_nq} except for the fact that positive values of $q$ are being used. It is clear that the difference between observed and theoretical value increases without limit as larger and larger positive values of q are being considered. Thus in \ref{Fig_L_M_pq} $q=0$ is the most favored model as this minimizes the difference between the observed and the theoretical luminosities when the entire sample is taken into consideration. 

\ref{Fig_L_dM_nq} , in contrast to \ref{Fig_L_M_nq}, depicts the variation of the calculated and the observed luminosities with respect to the mass accretion rates. In this case for intermediate and lower accretion rates, the difference between the observed and the calculated luminosities increases for more negative values of $q$. On the other hand, for higher accretion rates, this difference increases for less negative values of $q$. Thus, again we can presume that this difference will be minimized for the entire sample for some intermediate negative value of $q$. \ref{Fig_L_dM_pq} plots the same quantities as in \ref{Fig_L_dM_nq} except for positive values of $q$. The figure depicts that from intermediate to higher accretion rates the difference between the observed and the calculated luminosities increases with increasing $q$. Only for some quasars with low accretion rates this difference is more for lesser values of $q$. Thus, we can guess that $q=0$ is a more favored model in \ref{Fig_L_dM_pq}. 

\begin{figure}[t!]
\centering
\includegraphics[scale=1]{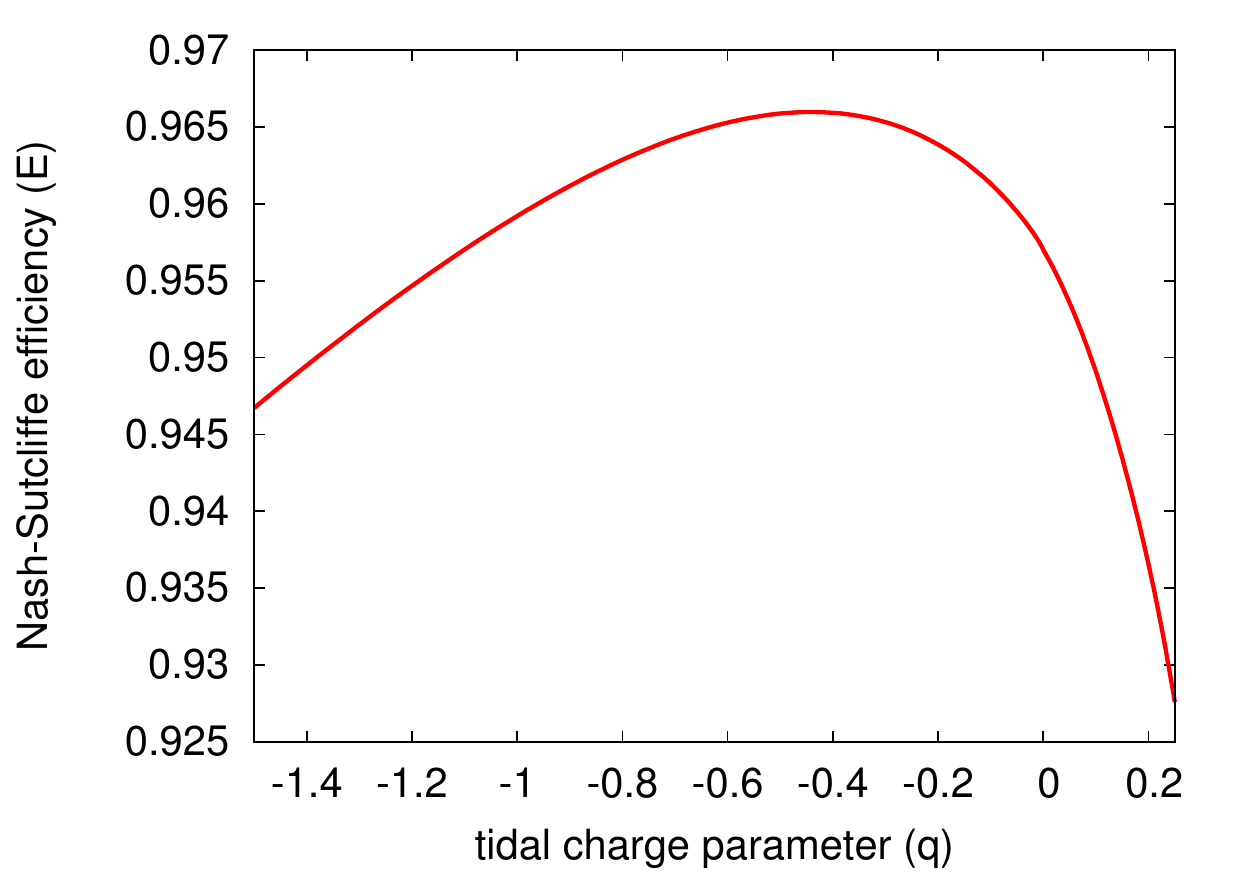}
\caption{The above figure depicts variation of the Nash-Sutcliffe efficiency with the tidal charge parameter $q$. In this case the maximum of $E$ minimizes the theoretical deviations from observational estimates. It is clear that once again negative value of $q$ maximizes the efficiency, in particular $E$ gets maximized at $q\sim -0.4$, close to the value which minimizes $\chi ^{2}_{\rm Red}$. }
\label{Fig_NSE}
\end{figure}

The above discussion qualitatively indicates that a negative value of the tidal charge parameter $q$ is more favored from observations. To achieve a more quantitative understanding of this result we briefly discuss several estimates of error along with their possible advantages and drawbacks. It is clear from the above discussion that for every value of $q$, for the entire sample of eighty quasars we have an observed and a theoretically estimated value corresponding to the optical luminosity. By comparing these values we wish to compute several estimates of error to find out the most favored value of $q$ and hence the observationally favored alternative gravity model.
\pagebreak
\begin{itemize}

\item \textbf{Reduced}~$\mathbf{\chi^{2}}~$:~

Given a set of observed data $\{\mathcal{O}_{i}\}$, with possible errors $\{\sigma_{i}\}$ and corresponding theoretical estimates $\{\mathcal{T}_{i}\}$ dependent on the charge parameter $q$, one can define the $\chi ^{2}$ distribution as,
\begin{align}\label{Acc_Eq_21}
\chi ^{2}(q)=\sum _{i}\frac{\left\{\mathcal{O}_{i}-\mathcal{T}_{i}(q) \right\}^{2}}{\sigma _{i}^{2}}
\end{align}
The errors in the above expression essentially provide proper weightage to each observations. Since in our case the errors $\sigma _{i}$ are not known, we simply consider each observation with equal weightage. This estimate, also known as the reduced $\chi ^{2}$ estimate will be a function of the tidal charge parameter $q$. The value of $q$ for which $\chi^{2}_{\rm Red}$ is minimized corresponds to the most favorable value of $q$. In \ref{Fig_chi}, we have plotted $\chi^{2}_{\rm Red}$ against the tidal charge parameter $q$. This explicitly demonstrates that as far as black hole continuum spectrum is considered, negative value of tidal charge parameter is more favored. This provides an indirect evidence for extra dimensions or scalar charges associated with black holes.

\begin{figure}[t!]
\centering
\includegraphics[scale=1]{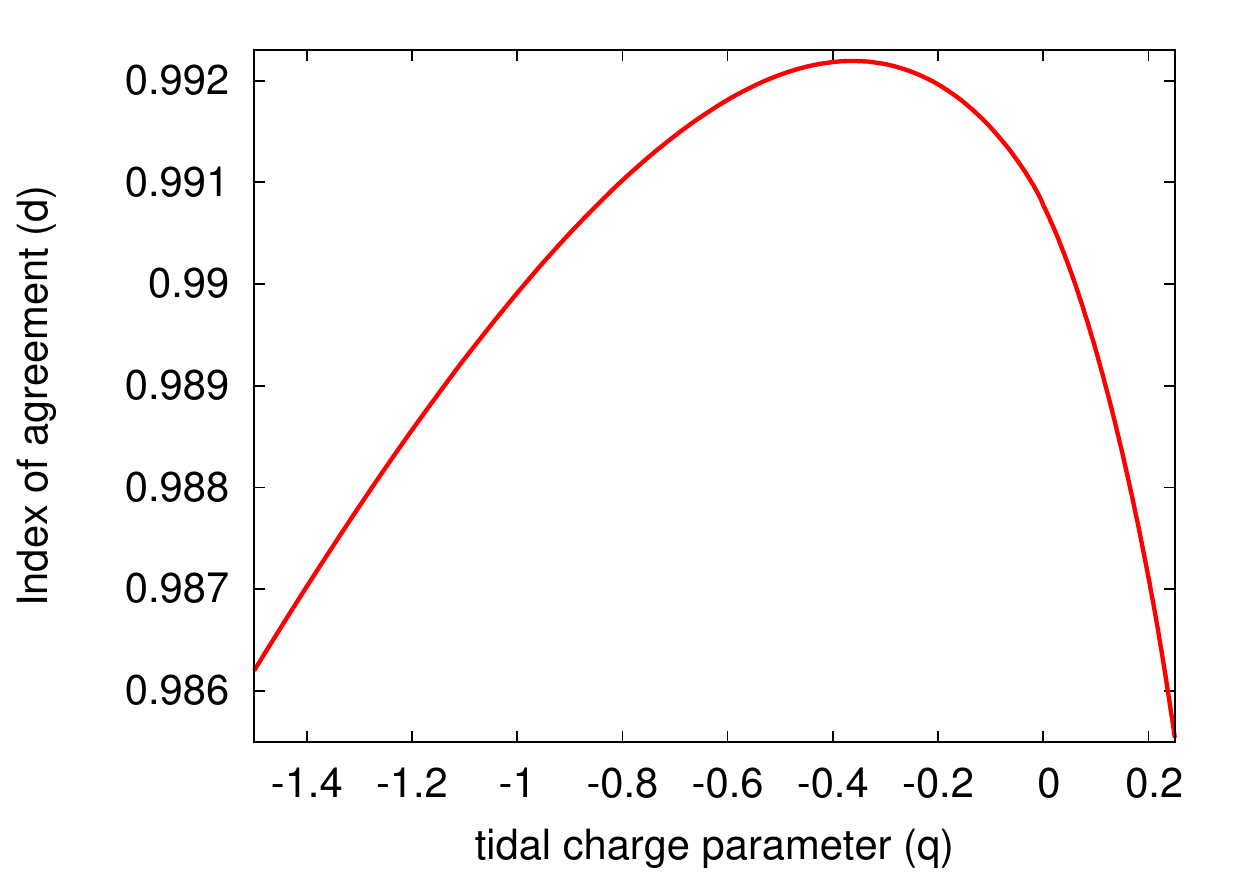}
\caption{Plot of index of agreement $d$ against the tidal charge parameter $q$. The plot shows that for negative values of $q$, the index of agreement is larger than for positive $q$ values. In particular for $q\sim -0.4$ it depicts a maximum, pointing to the fact that for this value of $q$, theoretical predictions are much closer to the observational one and thus it is more preferred compared to the Schwarzschild scenario. See text for more discussions.}
\label{Fig_Index}
\end{figure}

\item \textbf{Nash-Sutcliffe Efficiency:} This particular error estimator between theoretical predictions and observational results was first proposed by Nash and Sutcliffe \cite{NASH1970282,WRCR:WRCR8013,adgeo-5-89-2005}. This is again related to the sum of the absolute squared differences between the predicted and observed values, but normalized by the variance of the observed values in contrast to the $\chi^{2}$ estimator. This efficiency takes the following mathematical form,
\begin{align}\label{Acc_Eq_22}
E(q)=1-\frac{\sum_{i}\left\{\mathcal{O}_{i}-\mathcal{T}_{i}(q)\right\}^{2}}{\sum _{i}\left\{\mathcal{O}_{i}-\mathcal{O}_{\rm av}\right\}^{2}}
\end{align}
where $\mathcal{O}_{\rm av}$ stands for average observed value of the optical luminosity from the accreting disc. In contrast with the reduced $\chi ^{2}$ estimate, in this case, the maximum of the Nash-Sutcliffe efficiency depicts the value of tidal charge parameter minimizing the deviation of theoretical predictions from observational estimates. The variation of $E$ with tidal charge parameter $q$ has been presented in \ref{Fig_NSE}, where the maximization of Nash-Sutcliffe efficiency occurs for negative $q$ and more precisely for $q\sim -0.4$. Note that this is in very close agreement with the value of $q$, which minimizes $\chi^{2}_{\rm Red}$ as well. Hence once again the scenarios with higher dimensions or non-minimal scalar coupling to gravity and gauge fields is more preferred. 

\begin{figure}[t!]
\centering
\hbox{\hspace{-3em}
\includegraphics[scale=0.65]{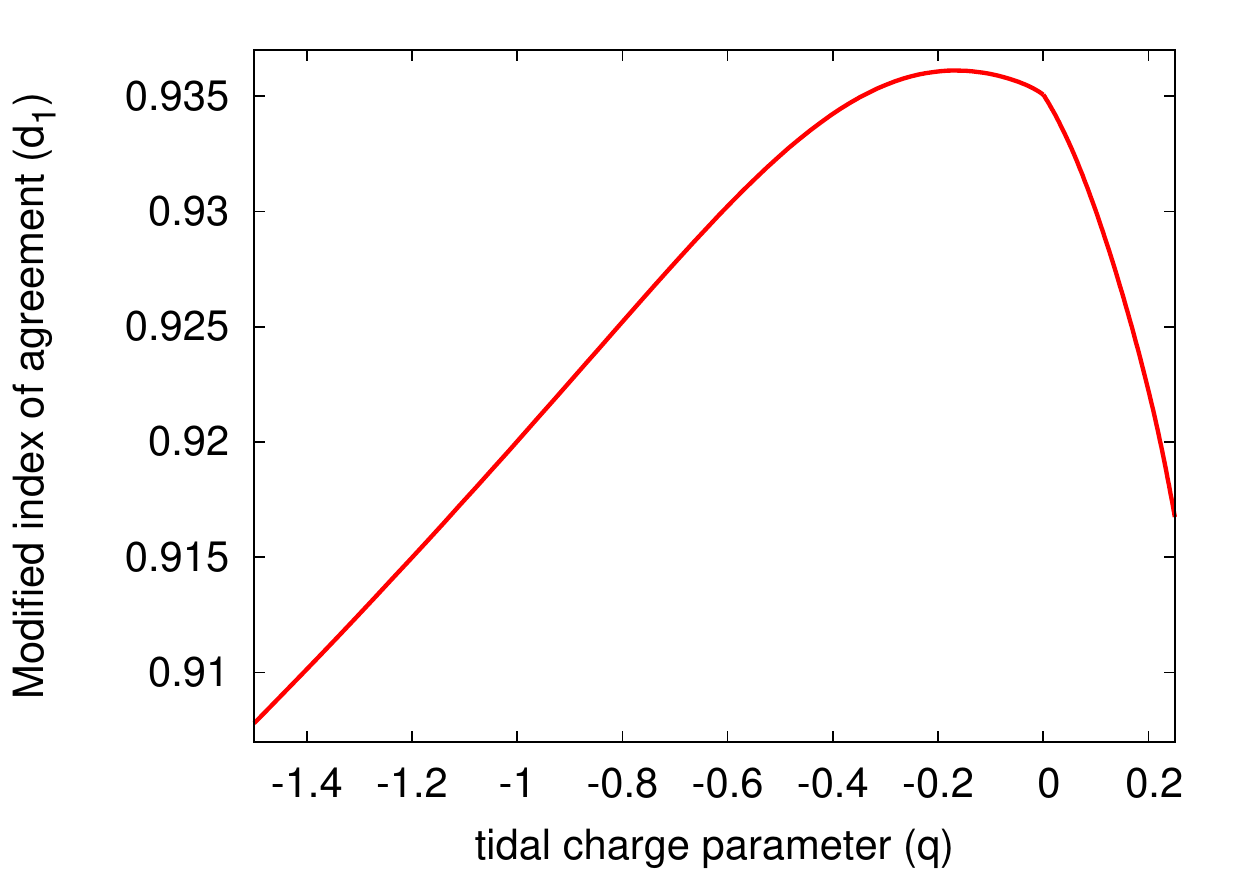}
\includegraphics[scale=0.65]{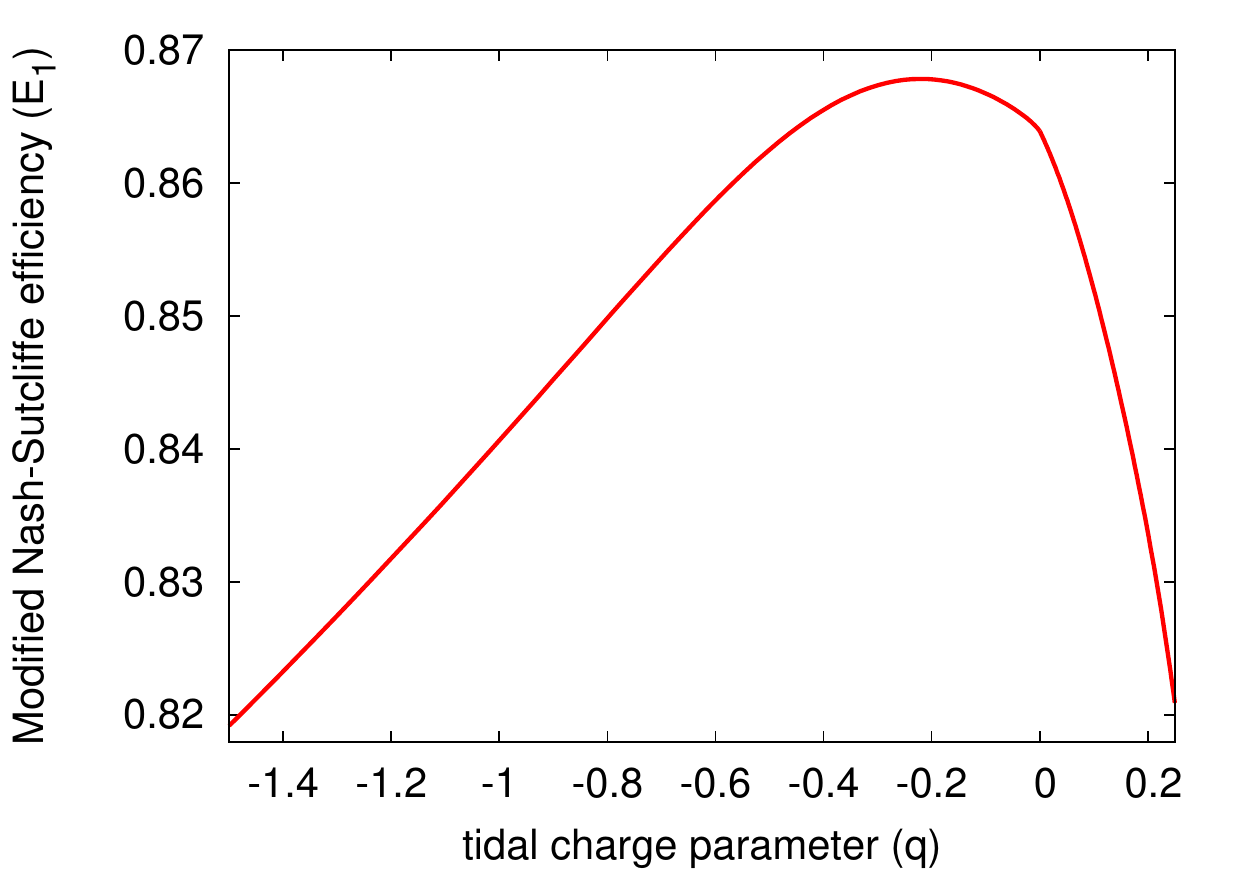}}
\caption{The above figures depicts variation of the modified error estimators with tidal charge parameter. The left one illustrates modified index of agreement $d_{1}$ while the right one corresponds to the modified Nash-Sutcliffe efficiency $E_{1}$. Both of them follows an identical trend of being maximum at some negative values of tidal charge parameter as fit with our earlier discussions.}
\label{Fig_Modified_Ed}
\end{figure}

\item \textbf{Index of agreement:} This error estimator is denoted as $d$ and was first proposed in \cite{Willmott1984,adgeo-5-89-2005} in order to overcome the insensitivity of Nash-Sutcliffe efficiency towards the differences between the observed and predicted means and variances \cite{WRCR:WRCR8013}. The above estimator is defined as:
\begin{align}\label{Acc_Eq_23}
d(q)=1-\frac{\sum_{i}\left\{\mathcal{O}_{i}-\mathcal{T}_{i}(q)\right\}^{2}}{\sum _{i}\left\{|\mathcal{O}_{i}-\mathcal{O}_{\rm av}|+|\mathcal{T}_{i}(q)-\mathcal{O}_{\rm av}|\right\}^{2}}
\end{align}
where again $\mathcal{O}_{\rm av}$ denotes average value of the observed luminosity. Since the index of agreement is a function of the tidal charge parameter $q$, its maximum value indicates the preferred value of $q$ leading to minimum deviation between theory and observations. Following the earlier trend, in this case the index of agreement peaks at $q\sim -0.4$ (see \ref{Fig_Index}), consistent with earlier observations and confirming possible signature of extra dimensions as well as scalar hairs.

\item \textbf{Modified Estimators:} One can use modified forms of the Nash-Sutcliffe efficiency as well as that of index of agreement   to overcome the oversensitivity of them to extreme observed values \cite{adgeo-5-89-2005}. This insensitivity is resulted by the mean square error in the Nash-Sutcliffe efficiency as well as in the index of agreement. In order to increase the sensitivity of these estimators for lower observational values a more general form of \ref{Acc_Eq_22} and \ref{Acc_Eq_23} can be put forward \cite{adgeo-5-89-2005}:
\begin{align}
E_{1}(q)&=1-\frac{\sum_{i}|\mathcal{O}_{i}-\mathcal{T}_{i}(q)|}{\sum _{i}|\mathcal{O}_{i}-\mathcal{O}_{\rm av}|}
\\
d_{1}(q)&=1-\frac{\sum_{i}|\mathcal{O}_{i}-\mathcal{T}_{i}(q)|}{\sum _{i}\left\{|\mathcal{O}_{i}-\mathcal{O}_{\rm av}|+|\mathcal{T}_{i}(q)-\mathcal{O}_{\rm av}| \right\}}
\end{align}
Both of these modified error estimators are dependent on the tidal charge parameter $q$ and again the most favorable value of $q$ must extremize both them. From \ref{Fig_Modified_Ed} it is clear that both of them goes through a maximum as the tidal charge parameter is being varied and the location of the maximum corresponds to $q\sim-0.3$, again negative and consistent with earlier error estimators. 

\end{itemize}
This shows that our earlier qualitative argument is revalidated by more quantitative error estimators as well, i.e., a negative value of $q$ is more favoured as far as black hole continuum spectrum is considered. To solidify the above claim a bit more, we will discuss about another potential source of uncertainty associated with this computation. This has to do with the choice of the inclination angle $i$, or in other words, the effect of the parameter $\cos i$ on the theoretical luminosity. In particular, we would like to discuss the stability of our claim in lieu of the various viable estimates of $\cos i$. Firstly, the unified model of the AGNs, discussed in \cite{Antonucci:1993sg}, indicates that the sample of the quasars considered in this work are more or less face on systems. This is further corroborated from the works of Polletta et al. \cite{Polletta:2007tf} and Reyes et al. \cite{Reyes:2008px}. Thus, we expect that the inclination angle should not exceed sixty degrees, i.e., maximum range of $\cos i \sim 0.5-1.0$, although a smaller range is expected. Further, $\cos i \sim 1$ is also ruled out because such systems are called blazars and in such cases the emission from the jet would be dominant compared to the disk emission. Secondly, the uncertainty in $\cos i$ is also going to affect the luminosity estimate in the Schwarzschild spacetime as well. To avoid this situation one can try to find out the value of $\cos i$ which may minimise the difference between observed and theoretical luminosities when the charge parameter $q=0$, i.e., in the Schwarzschild scenario itself. In this situation as well one can use all the machineries elaborated earlier, e.g., reduced $\chi^{2}$ test, Nash-Sutcliffe efficiency and so on, leading to the following range of values for $\cos i$: $0.77<\cos i <0.82$, minimising the error between theoretical and observed luminosities. Thus we have chosen a representative value, namely $\cos i=0.8$ in our work. Further, one can also check that for the inclination angles (or, $\cos i$) in the above range, if one introduces the charge parameter $q$, then again the error will be minimised for negative values of $q$ only. Thus if one minimises the error between the theoretical and observed luminosity one will end up with $\cos i \sim 0.8$ and negative values for the tidal charge parameter signalling possible existence of theories beyond \gr.

Before concluding this section, we would like to briefly mention about the effect of the tidal charge parameter on accretion rate and radiative efficiency respectively. For a particular numerical value of the tidal charge parameter $q$ one can determine the empirical relation between the accretion rate $\dot{M}$, the calculated optical luminosity $L_{\rm calc}$ and the black hole mass $M$. From such a empirical relation it is possible to comment on some features associated with the accretion rate. For example, in this particular situation one may infer that if the value of $q$ is being changed, the power law behaviour remains intact. Thus the tidal charge parameter has very little effect on the accretion rate. A similar empirical relation for the radiative efficiency $\eta$ can also be derived in terms of the black hole mass $M$ and theoretical luminosity $L_{\rm calc}$. In this case as well the empirical relation is insensitive to the value of the tidal charge parameter. Thus neither the accretion rate nor the radiative efficiency will differ considerably from the situation in \gr\ and hence will be consistent with the corresponding observations. 
\section{Concluding Remarks}\label{Accretion_Conc}

In this work our main aim was to understand the distinctive signature of alternative gravity theories in the context of black hole continuum spectrum. In particular we were mainly interested in possible effects on the electromagnetic emission from accretion disc around black holes due to presence of higher dimensions, higher curvature as well as scalar hairs in the theory. We have discussed three such possible scenarios --- (i) Higher dimensions manifested by modification of lower dimensional gravitational field equations, (ii) Einstein-Gauss-Bonnet gravity in higher spacetime dimensions modifying black hole solutions in four-dimensions and finally (iii) scalar hairs in black hole spacetime, inherited by Horndeski (or, generalised Horndeski) models. Interestingly in all the three scenarios, black hole solutions inherit tidal charge parameters which can be negative, in contrast with the \RN scenario. Thus if the sign of tidal charge parameter can be estimated by some means it will provide an interesting window to look for possible 
signatures of these alternative gravity models.

For this purpose, we have used the continuum spectrum emanating from accretion disc around supermassive black holes for this purpose, where the effect of gravity theory beyond Einstein is most prominent and the imprint of tidal charge parameter on the continuum spectrum can be estimated. Keeping this in mind, we have studied 80 quasars harboring supermassive black holes and have provided a qualitative estimate of the difference between theoretical and observational results, showing a minimization of error for negative values of the tidal charge parameter. We have made this statement more quantitative by discussing several estimators of errors --- (a) reduced $\chi^{2}$, (b) Nash-Sutcliffe Efficiency, (c) index of agreement and modified versions of the last two respectively. Most interestingly, all of them points towards a negative value for the tidal charge parameter, minimizing deviation from observations. This may indicate emergence of alternative gravity models harbouring black holes beyond Schwarzschild. 
The models with extra dimensions as well as scalar hairs are seemingly two such natural choices.

The above analysis opens up new observational avenues to explore, such as, measurement of strong gravitational lensing using VLBI, or understanding the gravitational waveform emanating from collision of two black holes with higher precision (e.g., using Advanced Laser Interferometer Gravitational-wave Observatory or Laser Interferometer Space Antenna) and their implications on these alternative gravity theories. It may also be worthwhile to explore the robustness of our result in more detail, e.g., whether the conclusion is affected if one considers some other sample of supermassive black holes or, how much dependent the results are on the statistics. Further, one can search for other alternative gravity theories, such as, gravity with torsion, pure Lovelock theories and $f(R)$ theories, in a similar context. We are currently pursuing some of these aspects and will possibly report elsewhere. 
\section*{Acknowledgements}

Research of S.C. is funded by the SERB-NPDF grant (PDF/2016/001589) from SERB, Government of India and the research of S.S.G is partially supported by the SERB-Extra Mural Research grant (EMR/2017/001372), Government of India. S.C. thanks Albert Einstein Institute, Golm, for warm hospitality where a part of this work was done.
\bibliography{Gravity_1_full,Gravity_2_full,Gravity_3_partial,Brane,My_References}

\providecommand{\href}[2]{#2}\begingroup\raggedright\begin{thebibliography}{100}

\bibitem{Clifton:2011jh}
T.~Clifton, P.~G. Ferreira, A.~Padilla, and C.~Skordis, ``{Modified Gravity and
  Cosmology},'' \href{http://dx.doi.org/10.1016/j.physrep.2012.01.001}{{\em
  Phys. Rept.} {\bfseries 513} (2012) 1--189},
\href{http://arxiv.org/abs/1106.2476}{{\ttfamily arXiv:1106.2476
  [astro-ph.CO]}}.

\bibitem{Perlmutter:1998np}
{\bfseries Supernova Cosmology Project} Collaboration, S.~Perlmutter {\em
  et~al.}, ``{Measurements of Omega and Lambda from 42 high redshift
  supernovae},'' \href{http://dx.doi.org/10.1086/307221}{{\em Astrophys. J.}
  {\bfseries 517} (1999) 565--586},
\href{http://arxiv.org/abs/astro-ph/9812133}{{\ttfamily arXiv:astro-ph/9812133
  [astro-ph]}}.

\bibitem{Riess:1998cb}
{\bfseries Supernova Search Team} Collaboration, A.~G. Riess {\em et~al.},
  ``{Observational evidence from supernovae for an accelerating universe and a
  cosmological constant},'' \href{http://dx.doi.org/10.1086/300499}{{\em
  Astron. J.} {\bfseries 116} (1998) 1009--1038},
\href{http://arxiv.org/abs/astro-ph/9805201}{{\ttfamily arXiv:astro-ph/9805201
  [astro-ph]}}.

\bibitem{Padmanabhan:2002ji}
T.~Padmanabhan, ``{Cosmological constant: The Weight of the vacuum},''
  \href{http://dx.doi.org/10.1016/S0370-1573(03)00120-0}{{\em Phys. Rept.}
  {\bfseries 380} (2003) 235--320},
\href{http://arxiv.org/abs/hep-th/0212290}{{\ttfamily arXiv:hep-th/0212290
  [hep-th]}}.

\bibitem{Copeland:2006wr}
E.~J. Copeland, M.~Sami, and S.~Tsujikawa, ``{Dynamics of dark energy},''
  \href{http://dx.doi.org/10.1142/S021827180600942X}{{\em Int. J. Mod. Phys.}
  {\bfseries D15} (2006) 1753--1936},
\href{http://arxiv.org/abs/hep-th/0603057}{{\ttfamily arXiv:hep-th/0603057
  [hep-th]}}.

\bibitem{Dadhich:2010ca}
N.~Dadhich, ``{On the enigmatic $\Lambda$ - a true constant of spacetime},''
  \href{http://dx.doi.org/10.1007/s12043-011-0163-7}{{\em Pramana} {\bfseries
  77} (2011) 433--437},
\href{http://arxiv.org/abs/1006.1552}{{\ttfamily arXiv:1006.1552 [gr-qc]}}.

\bibitem{Capozziello:2011et}
S.~Capozziello and M.~De~Laurentis, ``{Extended Theories of Gravity},''
  \href{http://dx.doi.org/10.1016/j.physrep.2011.09.003}{{\em Phys. Rept.}
  {\bfseries 509} (2011) 167--321},
\href{http://arxiv.org/abs/1108.6266}{{\ttfamily arXiv:1108.6266 [gr-qc]}}.

\bibitem{Nojiri:2017ncd}
S.~Nojiri, S.~D. Odintsov, and V.~K. Oikonomou, ``{Modified Gravity Theories on
  a Nutshell: Inflation, Bounce and Late-time Evolution},''
\href{http://arxiv.org/abs/1705.11098}{{\ttfamily arXiv:1705.11098 [gr-qc]}}.

\bibitem{Will:2014kxa}
C.~M. Will, ``{The Confrontation between General Relativity and Experiment},''
  \href{http://dx.doi.org/10.12942/lrr-2014-4}{{\em Living Rev. Rel.}
  {\bfseries 17} (2014) 4},
\href{http://arxiv.org/abs/1403.7377}{{\ttfamily arXiv:1403.7377 [gr-qc]}}.

\bibitem{Lanczos:1932zz}
C.~Lanczos, ``{Electricity as a natural property of Riemannian geometry},''
\href{http://dx.doi.org/10.1103/RevModPhys.39.716}{{\em Rev. Mod. Phys.}
  {\bfseries 39} (1932) 716--736}.

\bibitem{Lovelock:1971yv}
D.~Lovelock, ``{The Einstein tensor and its generalizations},''
\href{http://dx.doi.org/10.1063/1.1665613}{{\em J. Math. Phys.} {\bfseries 12}
  (1971) 498--501}.

\bibitem{gravitation}
T.Padmanabhan, {\em {Gravitation: Foundations and Frontiers}}.
\newblock Cambridge University Press, Cambridge, UK, 2010.

\bibitem{Padmanabhan:2013xyr}
T.~Padmanabhan and D.~Kothawala, ``{Lanczos-Lovelock models of gravity},''
  \href{http://dx.doi.org/10.1016/j.physrep.2013.05.007}{{\em Phys.Rept.}
  {\bfseries 531} (2013) 115--171},
\href{http://arxiv.org/abs/1302.2151}{{\ttfamily arXiv:1302.2151 [gr-qc]}}.

\bibitem{Dadhich:2008df}
N.~Dadhich, ``{Characterization of the Lovelock gravity by Bianchi
  derivative},'' \href{http://dx.doi.org/10.1007/s12043-010-0080-1}{{\em
  Pramana} {\bfseries 74} (2010) 875--882},
\href{http://arxiv.org/abs/0802.3034}{{\ttfamily arXiv:0802.3034 [gr-qc]}}.

\bibitem{Chakraborty:2014rga}
S.~Chakraborty and T.~Padmanabhan, ``{Evolution of Spacetime arises due to the
  departure from Holographic Equipartition in all Lanczos-Lovelock Theories of
  Gravity},'' \href{http://dx.doi.org/10.1103/PhysRevD.90.124017}{{\em
  Phys.Rev.} {\bfseries D90} no.~12, (2014) 124017},
\href{http://arxiv.org/abs/1408.4679}{{\ttfamily arXiv:1408.4679 [gr-qc]}}.

\bibitem{Chakraborty:2014joa}
S.~Chakraborty and T.~Padmanabhan, ``{Geometrical variables with direct
  thermodynamic significance in Lanczos-Lovelock gravity},''
  \href{http://dx.doi.org/10.1103/PhysRevD.90.084021}{{\em Phys.Rev.}
  {\bfseries D90} no.~8, (2014) 084021},
\href{http://arxiv.org/abs/1408.4791}{{\ttfamily arXiv:1408.4791 [gr-qc]}}.

\bibitem{Chakraborty:2015wma}
S.~Chakraborty, ``{Lanczos-Lovelock gravity from a thermodynamic
  perspective},'' \href{http://dx.doi.org/10.1007/JHEP08(2015)029}{{\em JHEP}
  {\bfseries 08} (2015) 029},
\href{http://arxiv.org/abs/1505.07272}{{\ttfamily arXiv:1505.07272 [gr-qc]}}.

\bibitem{Chakraborty:2015kva}
S.~Chakraborty and N.~Dadhich, ``{Brown-York quasilocal energy in
  Lanczos-Lovelock gravity and black hole horizons},''
  \href{http://dx.doi.org/10.1007/JHEP12(2015)003}{{\em JHEP} {\bfseries 12}
  (2015) 003},
\href{http://arxiv.org/abs/1509.02156}{{\ttfamily arXiv:1509.02156 [gr-qc]}}.

\bibitem{Lochan:2015bha}
K.~Lochan and S.~Chakraborty, ``{Discrete quantum spectrum of black holes},''
  \href{http://dx.doi.org/10.1016/j.physletb.2016.01.060}{{\em Phys. Lett.}
  {\bfseries B755} (2016) 37--42},
\href{http://arxiv.org/abs/1509.09010}{{\ttfamily arXiv:1509.09010 [gr-qc]}}.

\bibitem{Nojiri:2010wj}
S.~Nojiri and S.~D. Odintsov, ``{Unified cosmic history in modified gravity:
  from F(R) theory to Lorentz non-invariant models},''
  \href{http://dx.doi.org/10.1016/j.physrep.2011.04.001}{{\em Phys.Rept.}
  {\bfseries 505} (2011) 59--144},
\href{http://arxiv.org/abs/1011.0544}{{\ttfamily arXiv:1011.0544 [gr-qc]}}.

\bibitem{Nojiri:2003ft}
S.~Nojiri and S.~D. Odintsov, ``{Modified gravity with negative and positive
  powers of the curvature: Unification of the inflation and of the cosmic
  acceleration},'' \href{http://dx.doi.org/10.1103/PhysRevD.68.123512}{{\em
  Phys. Rev.} {\bfseries D68} (2003) 123512},
\href{http://arxiv.org/abs/hep-th/0307288}{{\ttfamily arXiv:hep-th/0307288
  [hep-th]}}.

\bibitem{Nojiri:2006gh}
S.~Nojiri and S.~D. Odintsov, ``{Modified f(R) gravity consistent with
  realistic cosmology: From matter dominated epoch to dark energy universe},''
  \href{http://dx.doi.org/10.1103/PhysRevD.74.086005}{{\em Phys. Rev.}
  {\bfseries D74} (2006) 086005},
\href{http://arxiv.org/abs/hep-th/0608008}{{\ttfamily arXiv:hep-th/0608008
  [hep-th]}}.

\bibitem{Capozziello:2006dj}
S.~Capozziello, S.~Nojiri, S.~D. Odintsov, and A.~Troisi, ``{Cosmological
  viability of f(R)-gravity as an ideal fluid and its compatibility with a
  matter dominated phase},''
  \href{http://dx.doi.org/10.1016/j.physletb.2006.06.034}{{\em Phys. Lett.}
  {\bfseries B639} (2006) 135--143},
\href{http://arxiv.org/abs/astro-ph/0604431}{{\ttfamily arXiv:astro-ph/0604431
  [astro-ph]}}.

\bibitem{Bahamonde:2016wmz}
S.~Bahamonde, S.~D. Odintsov, V.~K. Oikonomou, and M.~Wright, ``{Correspondence
  of $F(R)$ Gravity Singularities in Jordan and Einstein Frames},''
\href{http://arxiv.org/abs/1603.05113}{{\ttfamily arXiv:1603.05113 [gr-qc]}}.

\bibitem{Chakraborty:2016ydo}
S.~Chakraborty and S.~SenGupta, ``{Solving higher curvature gravity
  theories},'' \href{http://dx.doi.org/10.1140/epjc/s10052-016-4394-0}{{\em
  Eur. Phys. J.} {\bfseries C76} no.~10, (2016) 552},
\href{http://arxiv.org/abs/1604.05301}{{\ttfamily arXiv:1604.05301 [gr-qc]}}.

\bibitem{Shiromizu:1999wj}
T.~Shiromizu, K.-i. Maeda, and M.~Sasaki, ``{The Einstein equation on the
  3-brane world},'' \href{http://dx.doi.org/10.1103/PhysRevD.62.024012}{{\em
  Phys.Rev.} {\bfseries D62} (2000) 024012},
\href{http://arxiv.org/abs/gr-qc/9910076}{{\ttfamily arXiv:gr-qc/9910076
  [gr-qc]}}.

\bibitem{Dadhich:2000am}
N.~Dadhich, R.~Maartens, P.~Papadopoulos, and V.~Rezania, ``{Black holes on the
  brane},'' \href{http://dx.doi.org/10.1016/S0370-2693(00)00798-X}{{\em
  Phys.Lett.} {\bfseries B487} (2000) 1--6},
\href{http://arxiv.org/abs/hep-th/0003061}{{\ttfamily arXiv:hep-th/0003061
  [hep-th]}}.

\bibitem{Harko:2004ui}
T.~Harko and M.~Mak, ``{Vacuum solutions of the gravitational field equations
  in the brane world model},''
  \href{http://dx.doi.org/10.1103/PhysRevD.69.064020}{{\em Phys.Rev.}
  {\bfseries D69} (2004) 064020},
\href{http://arxiv.org/abs/gr-qc/0401049}{{\ttfamily arXiv:gr-qc/0401049
  [gr-qc]}}.

\bibitem{Carames:2012gr}
T.~R.~P. Carames, M.~E.~X. Guimaraes, and J.~M. Hoff~da Silva, ``{Effective
  gravitational equations for $f(R)$ braneworld models},''
  \href{http://dx.doi.org/10.1103/PhysRevD.87.106011}{{\em Phys. Rev.}
  {\bfseries D87} no.~10, (2013) 106011},
\href{http://arxiv.org/abs/1205.4980}{{\ttfamily arXiv:1205.4980 [gr-qc]}}.

\bibitem{Kobayashi:2006jw}
T.~Kobayashi, T.~Shiromizu, and N.~Deruelle, ``{Low energy effective
  gravitational equations on a Gauss-Bonnet brane},''
  \href{http://dx.doi.org/10.1103/PhysRevD.74.104031}{{\em Phys. Rev.}
  {\bfseries D74} (2006) 104031},
\href{http://arxiv.org/abs/hep-th/0608166}{{\ttfamily arXiv:hep-th/0608166
  [hep-th]}}.

\bibitem{Shiromizu:2002qr}
T.~Shiromizu and K.~Koyama, ``{Low-energy effective theory for two brane
  systems: Covariant curvature formulation},''
  \href{http://dx.doi.org/10.1103/PhysRevD.67.084022}{{\em Phys. Rev.}
  {\bfseries D67} (2003) 084022},
\href{http://arxiv.org/abs/hep-th/0210066}{{\ttfamily arXiv:hep-th/0210066
  [hep-th]}}.

\bibitem{Haghani:2012zq}
Z.~Haghani, H.~R. Sepangi, and S.~Shahidi, ``{Cosmological dynamics of brane
  f(R) gravity},'' \href{http://dx.doi.org/10.1088/1475-7516/2012/02/031}{{\em
  JCAP} {\bfseries 1202} (2012) 031},
\href{http://arxiv.org/abs/1201.6448}{{\ttfamily arXiv:1201.6448 [gr-qc]}}.

\bibitem{Borzou:2009gn}
A.~Borzou, H.~R. Sepangi, S.~Shahidi, and R.~Yousefi, ``{Brane f(R) gravity},''
  \href{http://dx.doi.org/10.1209/0295-5075/88/29001}{{\em Europhys. Lett.}
  {\bfseries 88} (2009) 29001},
\href{http://arxiv.org/abs/0910.1933}{{\ttfamily arXiv:0910.1933 [gr-qc]}}.

\bibitem{Chakraborty:2014xla}
S.~Chakraborty and S.~SenGupta, ``{Spherically symmetric brane spacetime with
  bulk $f(\mathcal {R})$ gravity},''
  \href{http://dx.doi.org/10.1140/epjc/s10052-014-3234-3}{{\em Eur.Phys.J.}
  {\bfseries C75} no.~1, (2015) 11},
\href{http://arxiv.org/abs/1409.4115}{{\ttfamily arXiv:1409.4115 [gr-qc]}}.

\bibitem{Chakraborty:2015bja}
S.~Chakraborty and S.~SenGupta, ``{Effective gravitational field equations on
  $m$-brane embedded in n-dimensional bulk of Einstein and $f(\mathcal {R})$
  gravity},'' \href{http://dx.doi.org/10.1140/epjc/s10052-015-3768-z}{{\em Eur.
  Phys. J.} {\bfseries C75} no.~11, (2015) 538},
\href{http://arxiv.org/abs/1504.07519}{{\ttfamily arXiv:1504.07519 [gr-qc]}}.

\bibitem{Chakraborty:2015taq}
S.~Chakraborty and S.~SenGupta, ``{Spherically symmetric brane in a bulk of
  f(R) and Gauss-Bonnet Gravity},''
  \href{http://dx.doi.org/10.1088/0264-9381/33/22/225001}{{\em Class. Quant.
  Grav.} {\bfseries 33} no.~22, (2016) 225001},
\href{http://arxiv.org/abs/1510.01953}{{\ttfamily arXiv:1510.01953 [gr-qc]}}.

\bibitem{Chakraborty:2016gpg}
S.~Chakraborty and S.~SenGupta, ``{Gravity stabilizes itself},''
\href{http://arxiv.org/abs/1701.01032}{{\ttfamily arXiv:1701.01032 [gr-qc]}}.

\bibitem{Horndeski:1974wa}
G.~W. Horndeski, ``{Second-order scalar-tensor field equations in a
  four-dimensional space},''
\href{http://dx.doi.org/10.1007/BF01807638}{{\em Int. J. Theor. Phys.}
  {\bfseries 10} (1974) 363--384}.

\bibitem{Sotiriou:2013qea}
T.~P. Sotiriou and S.-Y. Zhou, ``{Black hole hair in generalized scalar-tensor
  gravity},'' \href{http://dx.doi.org/10.1103/PhysRevLett.112.251102}{{\em
  Phys. Rev. Lett.} {\bfseries 112} (2014) 251102},
\href{http://arxiv.org/abs/1312.3622}{{\ttfamily arXiv:1312.3622 [gr-qc]}}.

\bibitem{Babichev:2016rlq}
E.~Babichev, C.~Charmousis, and A.~Lehébel, ``{Black holes and stars in
  Horndeski theory},''
  \href{http://dx.doi.org/10.1088/0264-9381/33/15/154002}{{\em Class. Quant.
  Grav.} {\bfseries 33} no.~15, (2016) 154002},
\href{http://arxiv.org/abs/1604.06402}{{\ttfamily arXiv:1604.06402 [gr-qc]}}.

\bibitem{Brans:1961sx}
C.~Brans and R.~H. Dicke, ``{Mach's principle and a relativistic theory of
  gravitation},''
\href{http://dx.doi.org/10.1103/PhysRev.124.925}{{\em Phys. Rev.} {\bfseries
  124} (1961) 925--935}.

\bibitem{Herdeiro:2015waa}
C.~A.~R. Herdeiro and E.~Radu, ``{Asymptotically flat black holes with scalar
  hair: a review},'' \href{http://dx.doi.org/10.1142/S0218271815420146}{{\em
  Int. J. Mod. Phys.} {\bfseries D24} no.~09, (2015) 1542014},
\href{http://arxiv.org/abs/1504.08209}{{\ttfamily arXiv:1504.08209 [gr-qc]}}.

\bibitem{Gleyzes:2014dya}
J.~Gleyzes, D.~Langlois, F.~Piazza, and F.~Vernizzi, ``{Healthy theories beyond
  Horndeski},'' \href{http://dx.doi.org/10.1103/PhysRevLett.114.211101}{{\em
  Phys. Rev. Lett.} {\bfseries 114} no.~21, (2015) 211101},
\href{http://arxiv.org/abs/1404.6495}{{\ttfamily arXiv:1404.6495 [hep-th]}}.

\bibitem{Langlois:2015skt}
D.~Langlois and K.~Noui, ``{Hamiltonian analysis of higher derivative
  scalar-tensor theories},''
  \href{http://dx.doi.org/10.1088/1475-7516/2016/07/016}{{\em JCAP} {\bfseries
  1607} no.~07, (2016) 016},
\href{http://arxiv.org/abs/1512.06820}{{\ttfamily arXiv:1512.06820 [gr-qc]}}.

\bibitem{Deffayet:2009wt}
C.~Deffayet, G.~Esposito-Farese, and A.~Vikman, ``{Covariant Galileon},''
  \href{http://dx.doi.org/10.1103/PhysRevD.79.084003}{{\em Phys. Rev.}
  {\bfseries D79} (2009) 084003},
\href{http://arxiv.org/abs/0901.1314}{{\ttfamily arXiv:0901.1314 [hep-th]}}.

\bibitem{VanAcoleyen:2011mj}
K.~Van~Acoleyen and J.~Van~Doorsselaere, ``{Galileons from Lovelock actions},''
  \href{http://dx.doi.org/10.1103/PhysRevD.83.084025}{{\em Phys. Rev.}
  {\bfseries D83} (2011) 084025},
\href{http://arxiv.org/abs/1102.0487}{{\ttfamily arXiv:1102.0487 [gr-qc]}}.

\bibitem{MuellerHoissen:1989yv}
F.~Mueller-Hoissen, ``{Gravity Actions, Boundary Terms and Second Order Field
  Equations},''
\href{http://dx.doi.org/10.1016/0550-3213(90)90513-D}{{\em Nucl. Phys.}
  {\bfseries B337} (1990) 709--736}.

\bibitem{Parattu:2016trq}
K.~Parattu, S.~Chakraborty, and T.~Padmanabhan, ``{Variational Principle for
  Gravity with Null and Non-null boundaries: A Unified Boundary
  Counter-term},'' \href{http://dx.doi.org/10.1140/epjc/s10052-016-3979-y}{{\em
  Eur. Phys. J.} {\bfseries C76} no.~3, (2016) 129},
\href{http://arxiv.org/abs/1602.07546}{{\ttfamily arXiv:1602.07546 [gr-qc]}}.

\bibitem{Babichev:2012re}
E.~Babichev and G.~Esposito-Farèse, ``{Time-Dependent Spherically Symmetric
  Covariant Galileons},''
  \href{http://dx.doi.org/10.1103/PhysRevD.87.044032}{{\em Phys. Rev.}
  {\bfseries D87} (2013) 044032},
\href{http://arxiv.org/abs/1212.1394}{{\ttfamily arXiv:1212.1394 [gr-qc]}}.

\bibitem{Kobayashi:2014eva}
T.~Kobayashi and N.~Tanahashi, ``{Exact black hole solutions in shift symmetric
  scalar–tensor theories},''
  \href{http://dx.doi.org/10.1093/ptep/ptu096}{{\em PTEP} {\bfseries 2014}
  (2014) 073E02},
\href{http://arxiv.org/abs/1403.4364}{{\ttfamily arXiv:1403.4364 [gr-qc]}}.

\bibitem{Babichev:2015rva}
E.~Babichev, C.~Charmousis, and M.~Hassaine, ``{Charged Galileon black
  holes},'' \href{http://dx.doi.org/10.1088/1475-7516/2015/05/031}{{\em JCAP}
  {\bfseries 1505} (2015) 031},
\href{http://arxiv.org/abs/1503.02545}{{\ttfamily arXiv:1503.02545 [gr-qc]}}.

\bibitem{Charmousis:2015txa}
C.~Charmousis and M.~Tsoukalas, ``{Lovelock Galileons and black holes},''
  \href{http://dx.doi.org/10.1103/PhysRevD.92.104050}{{\em Phys. Rev.}
  {\bfseries D92} no.~10, (2015) 104050},
\href{http://arxiv.org/abs/1506.05014}{{\ttfamily arXiv:1506.05014 [gr-qc]}}.

\bibitem{Bhattacharya:2016naa}
S.~Bhattacharya and S.~Chakraborty, ``{Constraining some Horndeski gravity
  theories},'' \href{http://dx.doi.org/10.1103/PhysRevD.95.044037}{{\em Phys.
  Rev.} {\bfseries D95} no.~4, (2017) 044037},
\href{http://arxiv.org/abs/1607.03693}{{\ttfamily arXiv:1607.03693 [gr-qc]}}.

\bibitem{Anabalon:2013oea}
A.~Anabalon, A.~Cisterna, and J.~Oliva, ``{Asymptotically locally AdS and flat
  black holes in Horndeski theory},''
  \href{http://dx.doi.org/10.1103/PhysRevD.89.084050}{{\em Phys. Rev.}
  {\bfseries D89} (2014) 084050},
\href{http://arxiv.org/abs/1312.3597}{{\ttfamily arXiv:1312.3597 [gr-qc]}}.

\bibitem{Cisterna:2014nua}
A.~Cisterna and C.~Erices, ``{Asymptotically locally AdS and flat black holes
  in the presence of an electric field in the Horndeski scenario},''
  \href{http://dx.doi.org/10.1103/PhysRevD.89.084038}{{\em Phys. Rev.}
  {\bfseries D89} (2014) 084038},
\href{http://arxiv.org/abs/1401.4479}{{\ttfamily arXiv:1401.4479 [gr-qc]}}.

\bibitem{Deffayet:2010zh}
C.~Deffayet, S.~Deser, and G.~Esposito-Farese, ``{Arbitrary $p$-form
  Galileons},'' \href{http://dx.doi.org/10.1103/PhysRevD.82.061501}{{\em Phys.
  Rev.} {\bfseries D82} (2010) 061501},
\href{http://arxiv.org/abs/1007.5278}{{\ttfamily arXiv:1007.5278 [gr-qc]}}.

\bibitem{Hinterbichler:2010xn}
K.~Hinterbichler, M.~Trodden, and D.~Wesley, ``{Multi-field galileons and
  higher co-dimension branes},''
  \href{http://dx.doi.org/10.1103/PhysRevD.82.124018}{{\em Phys. Rev.}
  {\bfseries D82} (2010) 124018},
\href{http://arxiv.org/abs/1008.1305}{{\ttfamily arXiv:1008.1305 [hep-th]}}.

\bibitem{Charmousis:2014mia}
C.~Charmousis, ``{From Lovelock to Horndeski`s Generalized Scalar Tensor
  Theory},'' \href{http://dx.doi.org/10.1007/978-3-319-10070-8_2}{{\em Lect.
  Notes Phys.} {\bfseries 892} (2015) 25--56},
\href{http://arxiv.org/abs/1405.1612}{{\ttfamily arXiv:1405.1612 [gr-qc]}}.

\bibitem{Virbhadra:1999nm}
K.~S. Virbhadra and G.~F.~R. Ellis, ``{Schwarzschild black hole lensing},''
  \href{http://dx.doi.org/10.1103/PhysRevD.62.084003}{{\em Phys. Rev.}
  {\bfseries D62} (2000) 084003},
\href{http://arxiv.org/abs/astro-ph/9904193}{{\ttfamily arXiv:astro-ph/9904193
  [astro-ph]}}.

\bibitem{Bozza:2001xd}
V.~Bozza, S.~Capozziello, G.~Iovane, and G.~Scarpetta, ``{Strong field limit of
  black hole gravitational lensing},''
  \href{http://dx.doi.org/10.1023/A:1012292927358}{{\em Gen. Rel. Grav.}
  {\bfseries 33} (2001) 1535--1548},
\href{http://arxiv.org/abs/gr-qc/0102068}{{\ttfamily arXiv:gr-qc/0102068
  [gr-qc]}}.

\bibitem{Bozza:2009yw}
V.~Bozza, ``{Gravitational Lensing by Black Holes},''
  \href{http://dx.doi.org/10.1007/s10714-010-0988-2}{{\em Gen. Rel. Grav.}
  {\bfseries 42} (2010) 2269--2300},
\href{http://arxiv.org/abs/0911.2187}{{\ttfamily arXiv:0911.2187 [gr-qc]}}.

\bibitem{Sahu:2015dea}
S.~Sahu, K.~Lochan, and D.~Narasimha, ``{Gravitational lensing by self-dual
  black holes in loop quantum gravity},''
  \href{http://dx.doi.org/10.1103/PhysRevD.91.063001}{{\em Phys. Rev.}
  {\bfseries D91} (2015) 063001},
\href{http://arxiv.org/abs/1502.05619}{{\ttfamily arXiv:1502.05619 [gr-qc]}}.

\bibitem{Bhadra:2003zs}
A.~Bhadra, ``{Gravitational lensing by a charged black hole of string
  theory},'' \href{http://dx.doi.org/10.1103/PhysRevD.67.103009}{{\em Phys.
  Rev.} {\bfseries D67} (2003) 103009},
\href{http://arxiv.org/abs/gr-qc/0306016}{{\ttfamily arXiv:gr-qc/0306016
  [gr-qc]}}.

\bibitem{Ghosh:2010uw}
T.~Ghosh and S.~Sengupta, ``{Strong gravitational lensing across dilaton
  anti-de Sitter black hole},''
  \href{http://dx.doi.org/10.1103/PhysRevD.81.044013}{{\em Phys. Rev.}
  {\bfseries D81} (2010) 044013},
\href{http://arxiv.org/abs/1001.5129}{{\ttfamily arXiv:1001.5129 [gr-qc]}}.

\bibitem{Ayzenberg:2017ufk}
D.~Ayzenberg and N.~Yunes, ``{Black Hole Continuum Spectra as a Test of General
  Relativity: Quadratic Gravity},''
  \href{http://dx.doi.org/10.1088/1361-6382/aa6dbc}{{\em Class. Quant. Grav.}
  {\bfseries 34} no.~11, (2017) 115003},
\href{http://arxiv.org/abs/1701.07003}{{\ttfamily arXiv:1701.07003 [gr-qc]}}.

\bibitem{Psaltis:2008bb}
D.~Psaltis, ``{Probes and Tests of Strong-Field Gravity with Observations in
  the Electromagnetic Spectrum},''
  \href{http://dx.doi.org/10.12942/lrr-2008-9}{{\em Living Rev. Rel.}
  {\bfseries 11} (2008) 9},
\href{http://arxiv.org/abs/0806.1531}{{\ttfamily arXiv:0806.1531 [astro-ph]}}.

\bibitem{Bambi:2015kza}
C.~Bambi, ``{Testing black hole candidates with electromagnetic radiation},''
  \href{http://dx.doi.org/10.1103/RevModPhys.89.025001}{{\em Rev. Mod. Phys.}
  {\bfseries 89} no.~2, (2017) 025001},
\href{http://arxiv.org/abs/1509.03884}{{\ttfamily arXiv:1509.03884 [gr-qc]}}.

\bibitem{Abramowicz:2011xu}
M.~A. Abramowicz and P.~C. Fragile, ``{Foundations of Black Hole Accretion Disk
  Theory},'' \href{http://dx.doi.org/10.12942/lrr-2013-1}{{\em Living Rev.
  Rel.} {\bfseries 16} (2013) 1},
\href{http://arxiv.org/abs/1104.5499}{{\ttfamily arXiv:1104.5499
  [astro-ph.HE]}}.

\bibitem{Zhang:2016ach}
{\bfseries eXTP} Collaboration, S.~N. Zhang {\em et~al.}, ``{eXTP -- enhanced
  X-ray Timing and Polarimetry Mission},''
  \href{http://dx.doi.org/10.1117/12.2232034}{{\em Proc. SPIE Int. Soc. Opt.
  Eng.} {\bfseries 9905} (2016) 99051Q},
\href{http://arxiv.org/abs/1607.08823}{{\ttfamily arXiv:1607.08823
  [astro-ph.IM]}}.

\bibitem{Fish:2009va}
V.~L. Fish and S.~S. Doeleman, ``{Observing a Black Hole Event Horizon:
  (Sub)Millimeter VLBI of Sgr A*},''
  \href{http://dx.doi.org/10.1017/S1743921309990500}{{\em IAU Symp.} {\bfseries
  261} (2010) 271--276},
\href{http://arxiv.org/abs/0906.4040}{{\ttfamily arXiv:0906.4040
  [astro-ph.GA]}}.

\bibitem{Jusufi:2017vew}
K.~Jusufi and A.~Övgün, ``{Light Deflection by a Quantum Improved Kerr Black
  Hole Pierced by a Cosmic String},''
\href{http://arxiv.org/abs/1707.02824}{{\ttfamily arXiv:1707.02824 [gr-qc]}}.

\bibitem{Shchigolev:2016gro}
V.~K. Shchigolev and D.~N. Bezbatko, ``{Studying Gravitational Deflection of
  Light by Kiselev Black Hole via Homotopy Perturbation Method},''
\href{http://arxiv.org/abs/1612.07279}{{\ttfamily arXiv:1612.07279 [gr-qc]}}.

\bibitem{Tsukamoto:2016jzh}
N.~Tsukamoto, ``{Deflection angle in the strong deflection limit in a general
  asymptotically flat, static, spherically symmetric spacetime},''
  \href{http://dx.doi.org/10.1103/PhysRevD.95.064035}{{\em Phys. Rev.}
  {\bfseries D95} no.~6, (2017) 064035},
\href{http://arxiv.org/abs/1612.08251}{{\ttfamily arXiv:1612.08251 [gr-qc]}}.

\bibitem{Andriot:2017oaz}
D.~Andriot and G.~Lucena~Gómez, ``{Signatures of extra dimensions in
  gravitational waves},''
  \href{http://dx.doi.org/10.1088/1475-7516/2017/06/048}{{\em JCAP} {\bfseries
  1706} no.~06, (2017) 048},
\href{http://arxiv.org/abs/1704.07392}{{\ttfamily arXiv:1704.07392 [hep-th]}}.

\bibitem{Chakraborty:2016lxo}
S.~Chakraborty and S.~SenGupta, ``{Strong gravitational lensing --- A probe for
  extra dimensions and Kalb-Ramond field},''
\href{http://arxiv.org/abs/1611.06936}{{\ttfamily arXiv:1611.06936 [gr-qc]}}.

\bibitem{LyndenBell:1969yx}
D.~Lynden-Bell, ``{Galactic nuclei as collapsed old quasars},''
\href{http://dx.doi.org/10.1038/223690a0}{{\em Nature} {\bfseries 223} (1969)
  690}.

\bibitem{Kazanas:2012sk}
D.~Kazanas, K.~Fukumura, E.~Behar, I.~Contopoulos, and C.~Shrader, ``{Toward a
  Unified AGN Structure},''
\href{http://arxiv.org/abs/1206.5022}{{\ttfamily arXiv:1206.5022
  [astro-ph.HE]}}.

\bibitem{Lynden_Rees:1971}
D.~{Lynden-Bell} and M.~J. {Rees}, ``{On quasars, dust and the galactic
  centre},'' \href{http://dx.doi.org/10.1093/mnras/152.4.461}{{\em Mon. Not.
  Roy. Astron. Soc.} {\bfseries 152} (1971) 461}.

\bibitem{Rees:1984si}
M.~J. Rees, ``{Black Hole Models for Active Galactic Nuclei},''
\href{http://dx.doi.org/10.1146/annurev.aa.22.090184.002351}{{\em Ann. Rev.
  Astron. Astrophys.} {\bfseries 22} (1984) 471--506}.

\bibitem{Sanders:1988rz}
D.~B. Sanders, B.~T. Soifer, J.~H. Elias, B.~F. Madore, K.~Matthews,
  G.~Neugebauer, and N.~Z. Scoville, ``{Ultraluminous infrared galaxies and the
  origin of quasars},''
\href{http://dx.doi.org/10.1086/165983}{{\em Astrophys. J.} {\bfseries 325}
  (1988) 74--91}.

\bibitem{Neugebauer:1987}
G.~{Neugebauer}, R.~F. {Green}, K.~{Matthews}, M.~{Schmidt}, B.~T. {Soifer},
  and J.~{Bennett}, ``{Continuum energy distributions of quasars in the
  Palomar-Green Survey},'' \href{http://dx.doi.org/10.1086/191175}{{\em
  Astrophys. J. Suppl.} {\bfseries 63} (Mar., 1987) 615--644}.

\bibitem{Baskin:2004wn}
A.~Baskin and A.~Laor, ``{What controls the C IV line profile in active
  galactic nuclei?},''
  \href{http://dx.doi.org/10.1111/j.1365-2966.2004.08525.x}{{\em Mon. Not. Roy.
  Astron. Soc.} {\bfseries 356} (2005) 1029--1044},
\href{http://arxiv.org/abs/astro-ph/0409196}{{\ttfamily arXiv:astro-ph/0409196
  [astro-ph]}}.

\bibitem{Scott:2004sv}
J.~E. Scott, G.~A. Kriss, M.~Brotherton, R.~F. Green, J.~Hutchings, J.~M.
  Shull, and W.~Zheng, ``{A Composite extreme ultraviolet QSO spectrum from
  FUSE},'' \href{http://dx.doi.org/10.1086/422336}{{\em Astrophys. J.}
  {\bfseries 615} (2004) 135--149},
\href{http://arxiv.org/abs/astro-ph/0407203}{{\ttfamily arXiv:astro-ph/0407203
  [astro-ph]}}.

\bibitem{Brandt:1999cm}
W.~N. Brandt, A.~Laor, and B.~J. Wills, ``{On the Nature of soft x-ray weak
  quasistellar objects},'' \href{http://dx.doi.org/10.1086/308207}{{\em
  Astrophys. J.} {\bfseries 528} (2000) 637--649},
\href{http://arxiv.org/abs/astro-ph/9908016}{{\ttfamily arXiv:astro-ph/9908016
  [astro-ph]}}.

\bibitem{Pringle_Rees:1972}
J.~E. {Pringle} and M.~J. {Rees}, ``{Accretion Disc Models for Compact X-Ray
  Sources},'' {\em Astron. and Astrophys.} {\bfseries 21} (Oct., 1972) 1.

\bibitem{Pringle:1981ds}
J.~E. Pringle, ``{Accretion discs in astrophysics},''
\href{http://dx.doi.org/10.1146/annurev.aa.19.090181.001033}{{\em Ann. Rev.
  Astron. Astrophys.} {\bfseries 19} (1981) 137--160}.

\bibitem{Shakura:1972te}
N.~I. Shakura and R.~A. Sunyaev, ``{Black holes in binary systems.
  Observational appearance},''
{\em Astron. Astrophys.} {\bfseries 24} (1973) 337--355.

\bibitem{Shakura:1976xk}
N.~I. Shakura and R.~A. Sunyaev, ``{A Theory of the instability of disk
  accretion on to black holes and the variability of binary X-ray sources,
  galactic nuclei and quasars},''
{\em Mon. Not. Roy. Astron. Soc.} {\bfseries 175} (1976) 613--632.

\bibitem{Abramowicz:1988sp}
M.~A. Abramowicz, B.~Czerny, J.~P. Lasota, and E.~Szuszkiewicz, ``{Slim
  accretion disks},''
\href{http://dx.doi.org/10.1086/166683}{{\em Astrophys. J.} {\bfseries 332}
  (1988) 646}.

\bibitem{Prendergast_Burbidge:1968}
K.~H. {Prendergast} and G.~R. {Burbidge}, ``{On the Nature of Some Galactic
  X-Ray Sources},'' {\em The Astrophysical Journal} {\bfseries 151} (Feb.,
  1968) L83.

\bibitem{Afshordi_2003}
N.~{Afshordi} and B.~{Paczy{\'n}ski}, ``{Geometrically Thin Disk Accreting into
  a Black Hole},'' \href{http://dx.doi.org/10.1086/375559}{{\em The
  Astrophysical Journal} {\bfseries 592} (July, 2003) 354--367},
  \href{http://arxiv.org/abs/astro-ph/0202409}{{\ttfamily astro-ph/0202409}}.

\bibitem{Frank:2002}
J.~Frank, A.~King, and D.~Raine, {\em {Accretion Power in Astrophysics}}.
\newblock Cambridge University Press, Cambridge, UK, 2002.

\bibitem{Sanders:1989tu}
D.~B. Sanders, E.~S. Phinney, G.~Neugebauer, B.~T. Soifer, and K.~Matthews,
  ``{Continuum energy distribution of quasars - Shapes and origins},''
\href{http://dx.doi.org/10.1086/168094}{{\em Astrophys. J.} {\bfseries 347}
  (1989) 29--51}.

\bibitem{Maccacaro:1991}
T.~{Maccacaro}, R.~{della Ceca}, I.~M. {Gioia}, S.~L. {Morris}, J.~T. {Stocke},
  and A.~{Wolter}, ``{The properties of X-ray-selected active galactic nuclei.
  I - Luminosity function, cosmological evolution, and contribution to the
  diffuse X-ray background},'' \href{http://dx.doi.org/10.1086/170102}{{\em
  Astrophys. J.} {\bfseries 374} (June, 1991) 117--133}.

\bibitem{dellaceca:1991}
R.~{della Ceca} and T.~{Maccacaro}, ``{The cosmological evolution and
  luminosity function of X-ray selected active galactic nuclei},'' in {\em The
  Space Distribution of Quasars}, D.~{Crampton}, ed., vol.~21 of {\em
  Astronomical Society of the Pacific Conference Series}, pp.~150--157.
\newblock 1991.

\bibitem{Reynolds:2013rva}
C.~S. Reynolds, ``{The Spin of Supermassive Black Holes},''
  \href{http://dx.doi.org/10.1088/0264-9381/30/24/244004}{{\em Class. Quant.
  Grav.} {\bfseries 30} (2013) 244004},
\href{http://arxiv.org/abs/1307.3246}{{\ttfamily arXiv:1307.3246
  [astro-ph.HE]}}.

\bibitem{Blackburne:2011}
J.~A. {Blackburne}, D.~{Pooley}, S.~{Rappaport}, and P.~L. {Schechter},
  ``{Sizes and Temperature Profiles of Quasar Accretion Disks from Chromatic
  Microlensing},'' \href{http://dx.doi.org/10.1088/0004-637X/729/1/34}{{\em
  Astrophys. J.} {\bfseries 729} (Mar., 2011) 34},
  \href{http://arxiv.org/abs/1007.1665}{{\ttfamily arXiv:1007.1665
  [astro-ph.CO]}}.

\bibitem{Jimenez-Vicente:2014lta}
J.~Jiménez-Vicente, E.~Mediavilla, C.~S. Kochanek, J.~A. Muñoz, V.~Motta,
  E.~Falco, and A.~M. Mosquera, ``{The Average Size and Temperature Profile of
  Quasar Accretion Disks},''
  \href{http://dx.doi.org/10.1088/0004-637X/783/1/47}{{\em Astrophys. J.}
  {\bfseries 783} (2014) 47},
\href{http://arxiv.org/abs/1401.2785}{{\ttfamily arXiv:1401.2785
  [astro-ph.CO]}}.

\bibitem{Thorne:1974ve}
K.~S. Thorne, ``{Disk accretion onto a black hole. 2. Evolution of the
  hole.},''
\href{http://dx.doi.org/10.1086/152991}{{\em Astrophys. J.} {\bfseries 191}
  (1974) 507--520}.

\bibitem{Page:1974he}
D.~N. Page and K.~S. Thorne, ``{Disk-Accretion onto a Black Hole. Time-Averaged
  Structure of Accretion Disk},''
\href{http://dx.doi.org/10.1086/152990}{{\em Astrophys. J.} {\bfseries 191}
  (1974) 499--506}.

\bibitem{Abramowicz_2010}
M.~A. {Abramowicz}, M.~{Jaroszy{\'n}ski}, S.~{Kato}, J.-P. {Lasota},
  A.~{R{\'o}{\.z}a{\'n}ska}, and A.~{S{\c a}dowski}, ``{Leaving the innermost
  stable circular orbit: the inner edge of a black-hole accretion disk at
  various luminosities},''
  \href{http://dx.doi.org/10.1051/0004-6361/201014467}{{\em Astron. and
  Astrophys.} {\bfseries 521} (Oct., 2010) A15},
  \href{http://arxiv.org/abs/1003.3887}{{\ttfamily arXiv:1003.3887
  [astro-ph.HE]}}.

\bibitem{Novikov_Thorne_1973}
I.~D. {Novikov} and K.~S. {Thorne}, ``{Astrophysics of black holes.},'' in {\em
  Black Holes (Les Astres Occlus)}, C.~{Dewitt} and B.~S. {Dewitt}, eds.,
  pp.~343--450.
\newblock 1973.

\bibitem{Riffert_Herold_1995}
H.~{Riffert} and H.~{Herold}, ``{Relativistic Accretion Disk Structure
  Revisited},'' \href{http://dx.doi.org/10.1086/176161}{{\em The Astrophysical
  Journal} {\bfseries 450} (Sept., 1995) 508}.

\bibitem{Davis:2010uq}
S.~W. Davis and A.~Laor, ``{The Radiative Efficiency of Accretion Flows in
  Individual AGN},'' \href{http://dx.doi.org/10.1088/0004-637X/728/2/98}{{\em
  Astrophys. J.} {\bfseries 728} (2011) 98},
\href{http://arxiv.org/abs/1012.3213}{{\ttfamily arXiv:1012.3213
  [astro-ph.CO]}}.

\bibitem{Schmidt:1983hr}
M.~Schmidt and R.~F. Green, ``{Quasar evolution derived from the Palomar bright
  quasar survey and other complete quasar surveys},''
\href{http://dx.doi.org/10.1086/161048}{{\em Astrophys. J.} {\bfseries 269}
  (1983) 352}.

\bibitem{Blaschke:2016uyo}
M.~Blaschke and Z.~Stuchlík, ``{Efficiency of the Keplerian accretion in
  braneworld Kerr-Newman spacetimes and mining instability of some naked
  singularity spacetimes},''
  \href{http://dx.doi.org/10.1103/PhysRevD.94.086006}{{\em Phys. Rev.}
  {\bfseries D94} no.~8, (2016) 086006},
\href{http://arxiv.org/abs/1610.07462}{{\ttfamily arXiv:1610.07462 [gr-qc]}}.

\bibitem{Stuchlik:2008fy}
Z.~Stuchlik and A.~Kotrlova, ``{Orbital resonances in discs around braneworld
  Kerr black holes},'' \href{http://dx.doi.org/10.1007/s10714-008-0709-2}{{\em
  Gen. Rel. Grav.} {\bfseries 41} (2009) 1305--1343},
\href{http://arxiv.org/abs/0812.5066}{{\ttfamily arXiv:0812.5066 [astro-ph]}}.

\bibitem{Schee:2008fc}
J.~Schee and Z.~Stuchlik, ``{Profiles of emission lines generated by rings
  orbiting braneworld Kerr black holes},''
  \href{http://dx.doi.org/10.1007/s10714-008-0753-y}{{\em Gen. Rel. Grav.}
  {\bfseries 41} (2009) 1795--1818},
\href{http://arxiv.org/abs/0812.3017}{{\ttfamily arXiv:0812.3017 [astro-ph]}}.

\bibitem{Boroson:1992cf}
T.~A. Boroson and R.~F. Green, ``{The Emission - line properties of low -
  redshift quasi-stellar objects},''
\href{http://dx.doi.org/10.1086/191661}{{\em Astrophys. J. Suppl.} {\bfseries
  80} (1992) 109}.

\bibitem{Ferrarese:2000se}
L.~Ferrarese and D.~Merritt, ``{A Fundamental relation between supermassive
  black holes and their host galaxies},''
  \href{http://dx.doi.org/10.1086/312838}{{\em Astrophys. J.} {\bfseries 539}
  (2000) L9},
\href{http://arxiv.org/abs/astro-ph/0006053}{{\ttfamily arXiv:astro-ph/0006053
  [astro-ph]}}.

\bibitem{Gebhardt:2000fk}
K.~Gebhardt {\em et~al.}, ``{A Relationship between nuclear black hole mass and
  galaxy velocity dispersion},'' \href{http://dx.doi.org/10.1086/312840}{{\em
  Astrophys. J.} {\bfseries 539} (2000) L13},
\href{http://arxiv.org/abs/astro-ph/0006289}{{\ttfamily arXiv:astro-ph/0006289
  [astro-ph]}}.

\bibitem{Dasyra:2006jz}
K.~M. Dasyra, L.~J. Tacconi, R.~I. Davies, R.~Genzel, D.~Lutz, B.~M. Peterson,
  S.~Veilleux, A.~J. Baker, M.~Schweitzer, and E.~Sturm, ``{Host dynamics and
  origin of Palomar-Green QSOs},'' \href{http://dx.doi.org/10.1086/510552}{{\em
  Astrophys. J.} {\bfseries 657} (2007) 102--115},
\href{http://arxiv.org/abs/astro-ph/0610719}{{\ttfamily arXiv:astro-ph/0610719
  [astro-ph]}}.

\bibitem{Wolf:2008sm}
M.~J. Wolf and A.~I. Sheinis, ``{Host Galaxies of Luminous Quasars: Structural
  Properties and the Fundamental Plane},''
  \href{http://dx.doi.org/10.1088/0004-6256/136/4/1587}{{\em Astron. J.}
  {\bfseries 136} (2008) 1587},
\href{http://arxiv.org/abs/0808.0918}{{\ttfamily arXiv:0808.0918 [astro-ph]}}.

\bibitem{Tremaine:2002js}
S.~Tremaine {\em et~al.}, ``{The slope of the black hole mass versus velocity
  dispersion correlation},'' \href{http://dx.doi.org/10.1086/341002}{{\em
  Astrophys. J.} {\bfseries 574} (2002) 740--753},
\href{http://arxiv.org/abs/astro-ph/0203468}{{\ttfamily arXiv:astro-ph/0203468
  [astro-ph]}}.

\bibitem{Boulware:1985wk}
D.~G. Boulware and S.~Deser, ``{String Generated Gravity Models},''
\href{http://dx.doi.org/10.1103/PhysRevLett.55.2656}{{\em Phys. Rev. Lett.}
  {\bfseries 55} (1985) 2656}.

\bibitem{Maeda:2006hj}
H.~Maeda and N.~Dadhich, ``{Matter without matter: Novel Kaluza-Klein spacetime
  in Einstein-Gauss-Bonnet gravity},''
  \href{http://dx.doi.org/10.1103/PhysRevD.75.044007}{{\em Phys. Rev.}
  {\bfseries D75} (2007) 044007},
\href{http://arxiv.org/abs/hep-th/0611188}{{\ttfamily arXiv:hep-th/0611188
  [hep-th]}}.

\bibitem{Babichev:2013cya}
E.~Babichev and C.~Charmousis, ``{Dressing a black hole with a time-dependent
  Galileon},'' \href{http://dx.doi.org/10.1007/JHEP08(2014)106}{{\em JHEP}
  {\bfseries 08} (2014) 106},
\href{http://arxiv.org/abs/1312.3204}{{\ttfamily arXiv:1312.3204 [gr-qc]}}.

\bibitem{Chakraborty:2015vla}
S.~Chakraborty, ``{Aspects of Neutrino Oscillation in Alternative Gravity
  Theories},'' \href{http://dx.doi.org/10.1088/1475-7516/2015/10/019}{{\em
  JCAP} {\bfseries 1510} no.~10, (2015) 019},
\href{http://arxiv.org/abs/1506.02647}{{\ttfamily arXiv:1506.02647 [gr-qc]}}.

\bibitem{NASH1970282}
J.~Nash and J.~Sutcliffe, ``River flow forecasting through conceptual models
  part i — a discussion of principles,''
  \href{http://dx.doi.org/http://dx.doi.org/10.1016/0022-1694(70)90255-6}{{\em
  Journal of Hydrology} {\bfseries 10} no.~3, (1970) 282 -- 290}.
  \url{http://www.sciencedirect.com/science/article/pii/0022169470902556}.

\bibitem{WRCR:WRCR8013}
D.~R. Legates and G.~J. McCabe, ``Evaluating the use of “goodness-of-fit”
  measures in hydrologic and hydroclimatic model validation,''
  \href{http://dx.doi.org/10.1029/1998WR900018}{{\em Water Resources Research}
  {\bfseries 35} no.~1, (1999) 233--241}.
  \url{http://dx.doi.org/10.1029/1998WR900018}.

\bibitem{adgeo-5-89-2005}
P.~Krause, D.~P. Boyle, and F.~B\"ase, ``Comparison of different efficiency
  criteria for hydrological model assessment,''
  \href{http://dx.doi.org/10.5194/adgeo-5-89-2005}{{\em Advances in
  Geosciences} {\bfseries 5} (2005) 89--97}.
  \url{https://www.adv-geosci.net/5/89/2005/}.

\bibitem{Willmott1984}
C.~J. Willmott, {\em On the Evaluation of Model Performance in Physical
  Geography},
  \href{http://dx.doi.org/10.1007/978-94-017-3048-8_23}{pp.~443--460}.
\newblock Springer Netherlands, Dordrecht, 1984.
\newblock \url{http://dx.doi.org/10.1007/978-94-017-3048-8_23}.

\bibitem{Antonucci:1993sg}
R.~Antonucci, ``{Unified models for active galactic nuclei and quasars},''
\href{http://dx.doi.org/10.1146/annurev.aa.31.090193.002353}{{\em Ann. Rev.
  Astron. Astrophys.} {\bfseries 31} (1993) 473--521}.

\bibitem{Polletta:2007tf}
M.~Polletta, D.~Weedman, S.~Hoenig, C.~J. Lonsdale, H.~E. Smith, and J.~Houck,
  ``{Obscuration in extremely luminous quasars},''
  \href{http://dx.doi.org/10.1086/524343}{{\em Astrophys. J.} {\bfseries 675}
  (2008) 960--984},
\href{http://arxiv.org/abs/0709.4458}{{\ttfamily arXiv:0709.4458 [astro-ph]}}.

\bibitem{Reyes:2008px}
R.~Reyes, N.~L. Zakamska, M.~A. Strauss, J.~Green, J.~H. Krolik, Y.~Shen,
  G.~Richards, S.~Anderson, and D.~Schneider, ``{Space Density of
  Optically-Selected Type 2 Quasars},''
  \href{http://dx.doi.org/10.1088/0004-6256/136/6/2373}{{\em Astron. J.}
  {\bfseries 136} (2008) 2373--2390},
\href{http://arxiv.org/abs/0801.1115}{{\ttfamily arXiv:0801.1115 [astro-ph]}}.

\end{thebibliography}\endgroup

\bibliographystyle{./utphys1}
\end{document}